\documentclass[%
reprint,
superscriptaddress,
bibnotes,
amsmath,amssymb,
aps,
floatfix
]{revtex4-1}

\usepackage[dvipdfmx]{graphicx}
\usepackage{graphicx}
\usepackage{amsmath,amssymb,amsthm,mathrsfs,amsfonts,dsfont}
\usepackage{subfigure, epsfig}
\usepackage{braket}
\usepackage{bm}
\usepackage{enumerate}
\usepackage{comment}
\usepackage[colorlinks,linkcolor=blue,citecolor=blue]{hyperref}
\newcommand{\tr}{\mathrm{Tr}}

\begin{document}
\title{Quantum Fluctuation Theorem under Continuous Measurement and Feedback}

\author{Toshihiro Yada}
\email{yada@noneq.t.u-tokyo.ac.jp}
\affiliation{Department of Applied Physics, University of Tokyo, 7-3-1 Hongo, Bunkyo-Ku, Tokyo 113-8656, Japan}

\author{Nobuyuki Yoshioka}
\affiliation{Department of Applied Physics, University of Tokyo, 7-3-1 Hongo, Bunkyo-Ku, Tokyo 113-8656, Japan}

\author{Takahiro Sagawa}
\affiliation{Department of Applied Physics, University of Tokyo, 7-3-1 Hongo, Bunkyo-Ku, Tokyo 113-8656, Japan}
\affiliation{Quantum-Phase Electronics Center (QPEC), The University of Tokyo, 7-3-1 Hongo, Bunkyo-ku, Tokyo 113-8656, Japan}


\begin{abstract}
While the fluctuation theorem in classical systems has been thoroughly generalized under various feedback control setups, an intriguing situation in quantum systems, namely under continuous feedback, remains to be investigated. In this work, we derive the generalized fluctuation theorem under continuous quantum measurement and feedback. The essence for the derivation is to newly introduce the operationally meaningful information, which we call \emph{quantum-classical-transfer} (\emph{QC-transfer}) \emph{entropy}. QC-transfer entropy can be naturally interpreted as the quantum counterpart of transfer entropy that is commonly used in classical time series analysis. We also verify our theoretical results by numerical simulation and propose an experiment-numerics hybrid verification method. Our work reveals a fundamental connection between quantum thermodynamics and quantum information, which can be experimentally tested with artificial quantum systems such as circuit quantum electrodynamics.
\end{abstract}
\maketitle

\emph{Introduction.---}In the last few decades, the framework of thermodynamics has been applied to small systems in which thermodynamic quantities behave stochastically due to the presence of thermal or quantum fluctuations \cite{seifert2012stochastic,RevModPhys.81.1665,campisi2011colloquium,funo2018quantum,PRXinfothermoExp}.
A key relation that quantifies a universal behavior of such systems is the fluctuation theorem (FT)
\begin{equation}
    \label{eq:fluc}\langle e^{-\sigma}\rangle =1,
\end{equation}
where \begin{math}\sigma\end{math} is the stochastic entropy production and $\langle \cdot \rangle$ denotes the ensemble average.
The FT characterizes the behavior of the entropy production even in the nonlinear nonequilibrium region, and also implies the second law of thermodynamics (SL) at the average level: $\langle \sigma \rangle \geq 0$.

\begin{figure}[tb]
\begin{center}
\includegraphics[width=0.45\textwidth]{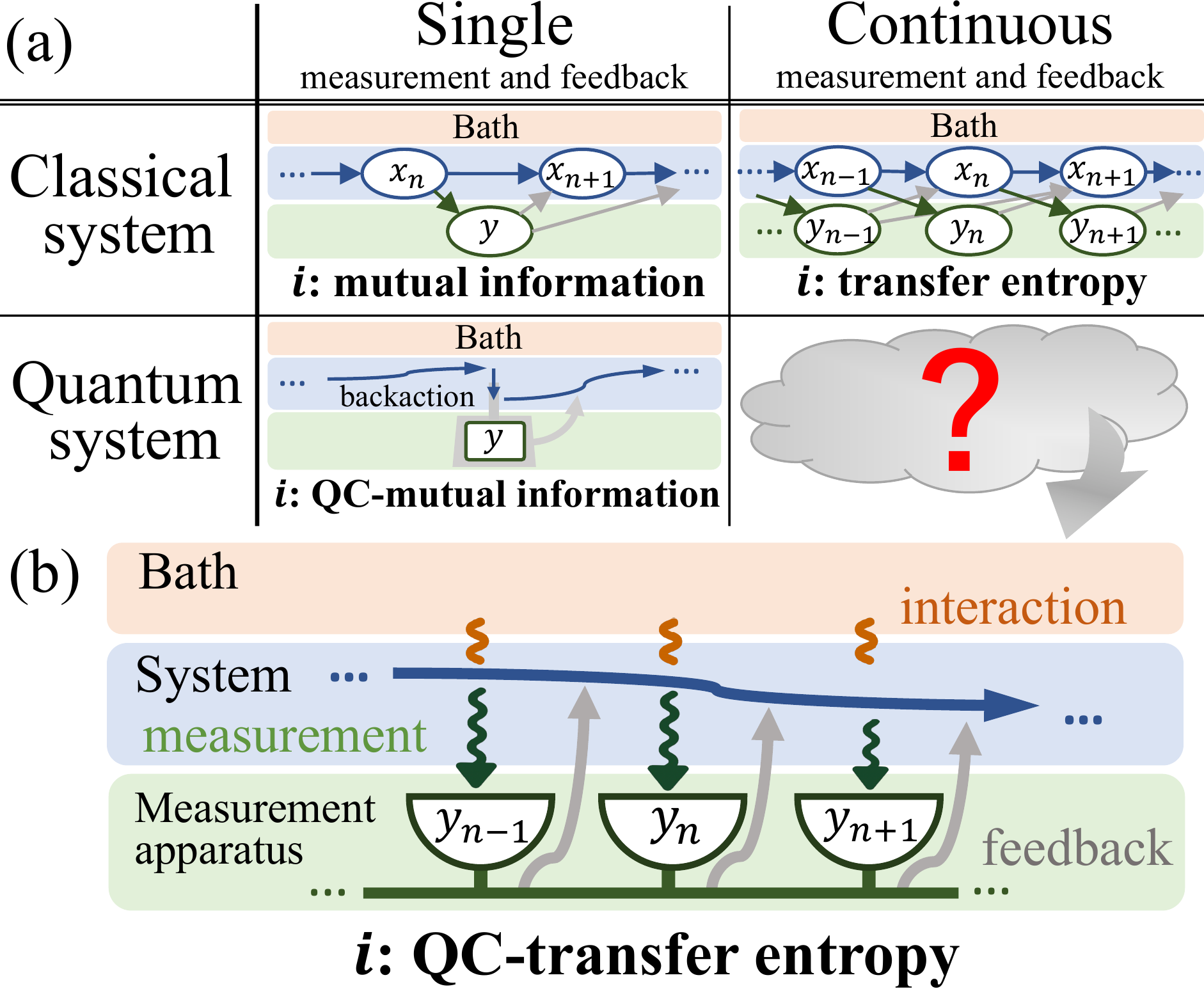}
\caption{(a) Summary of previous research of information thermodynamics. In all of the three cases shown here, the FT has been generalized in the form of Eq.~\eqref{intro_generalized FT} with appropriate choices of the information gain $i$, whereas quantum systems under continuous feedback control have not been studied. (b) Setup of the present work. We find that the QC-transfer entropy is the relevant information gain $i$ in this setup.}
\label{fig:setup}
\end{center}
\end{figure}

In light of thermodynamics of information, originated in the gedanken experiment of Maxwell's demon \cite{leff2002maxwell}, it has been revealed that measurement and feedback leads to generalizations of the laws of thermodynamics~\cite{parrondo2015thermodynamics,SagawaDoc,sagawa2010generalized,sagawa2012fluctuation,sagawa2012nonequilibrium,ito2013information,ito2016information,shiraishi2015fluctuation,HorowitzEspositoPRX2014,sagawa2008second,funo2013integral,gong2016quantum,toyabe2010experimental,roldan2014universal,koski2014experimental,vidrighin2016photonic,cottet2017observing,masuyama2018information,naghiloo2018information}. 
For instance, the FT has been generalized by incorporating information gain $i$ obtained from the measurement as
\begin{equation}
\label{intro_generalized FT}
\langle e^{-\sigma-i}\rangle =1,
\end{equation} 
which implies the generalized SL: \begin{math}\langle \sigma \rangle \geq -\langle i\rangle\end{math}.
The generalized FT in the form of Eq.~\eqref{intro_generalized FT} has been derived for classical systems under single measurement and feedback \cite{sagawa2010generalized,sagawa2012fluctuation} as well as continuous measurement and feedback ~\cite{sagawa2012nonequilibrium,ito2013information,ito2016information}, and also derived for quantum systems under single measurement and feedback \cite{funo2013integral,gong2016quantum} (see Fig.~\ref{fig:setup} (a)). There are also a few works about the role of continuous quantum measurement on the SL or the FT \cite{campisi2010fluctuation,belenchia2020entropy}.
However, the role of continuous measurement and \emph{feedback} in the quantum regime has not yet been elucidated, despite its significance as described below.

Continuous measurement and feedback has been of keen interest due to its capability of creating and stabilizing desired quantum states via feedback loop.
It is relevant to various quantum systems~\cite{wiseman2009quantum} including its applications to thermodynamics \cite{viisanen2015incomplete,karimi2020quantum}, and is developing due to the recent advancement of experimental techniques \cite{guerlin2007progressive,deleglise2008reconstruction,hofmann2016measuring,kurzmann2019optical,minev2019catch,gasparinetti2015fast,karimi2020reaching}.
It is also noteworthy that arbitrary Markovian open quantum systems described by the Lindblad master equation can be interpreted to be under continuous (non-selective) measurement \cite{wiseman2009quantum}. These facts tells us the framework of continuous quantum measurement and feedback gives a unified description of systems under artificial control or interaction with external systems.

In this Letter, we generalize the SL and the FT for quantum systems under continuous measurement and feedback, by introducing the \emph{quantum-classical-transfer} (\emph{QC-transfer}) \emph{entropy} as a relevant information gain.
The QC-transfer entropy is defined as the accumulation over time of the conditional QC-mutual information \cite{groenewold1971problem,ozawa1986information,buscemi2008global,sagawa2013second} under the past measurement outcomes. Therefore, it can naturally be interpreted as the quantum counterpart of the transfer entropy \cite{schreiber2000measuring} that is used to derive the generalized FT for classical systems~\cite{sagawa2012nonequilibrium,ito2013information,ito2016information}.
We also verify the generalized FT by numerical simulation in a two-level system, and propose an experiment-numerics hybrid verification method of the generalized FT.

\emph{Dynamics of the system.---}
Let us consider a quantum system interacting with the heat bath at inverse temperature $\beta$ under continuous measurement and feedback (Fig.~\ref{fig:setup}(b)). 
To simplify the argument, we suppose that the Born-Markov and rotating-wave approximations can be applied to the system-bath interaction \cite{breuer2002theory,albash2012quantum}.
We discretize time as $t_n\equiv n \Delta t$, consider the time evolution from $t=0$ to $t=\tau\equiv t_N$, and later take the continuous time limit $\Delta t \to 0, N\to \infty$ while keeping $\tau$ constant. The time evolution in $[t_n,t_{n+1})$ is described by the stochastic master equation:
\begin{equation}
    \label{SME}
    \begin{split}
    \rho_{t_{n+1}}^{Y_{n+1}} & = \rho_{t_{n}}^{Y_{n}}+\sum_{y} \Delta N_{y} \mathcal{G} [M_{y}] \rho_{t_n}^{Y_n} \\
    +&\Delta t\Bigl\{-i[H_{t_n}+h_{t_n},\rho_{t_n}^{Y_n}] + \sum_{d}\mathcal{D}[L_{d}]\rho_{t_n}^{Y_n} \\
    +& \sum_{y}-\frac{1}{2}\{M_{y}^\dag M_{y},\rho_{t_n}^{Y_n}\} + \tr[M_{y}\rho_{t_n}^{Y_n}M_{y}^\dag]\rho_{t_n}^{Y_n}\Bigr\},
    \end{split}
\end{equation}
where \begin{math}\mathcal{G}[m]\rho \equiv (m\rho m^\dag /\tr[ m\rho m^\dag]) -\rho \end{math} and \begin{math} \mathcal{D}[c]\rho\equiv c\rho c^\dag -1/2\{c^\dag c,\rho\}\end{math}. 
We define $y_n$ as the newly obtained measurement result at $t_n$ and $Y_n\equiv(y_1,y_2\dots ,y_n)$ as the outcomes until $t_n$. Here, $\rho_{t_n}^{Y_n}$ represents the conditional density operator at $t_n$ when the measurement results are \begin{math}Y_n\end{math}. 
We define $H_{t}$ as the intrinsic system Hamiltonian and $h_{t}$ as the external driving Hamiltonian.
The interaction between the system and the heat bath can be described by the Lindblad operators $\{L_{d}\}$, where $L_{d}$ represents the dissipation of energy $\Delta_{d}$ to the heat bath (i.e., $[L_{d},H_{t_n}]=\Delta_{d}L_{d}$), and satisfies the detailed balance condition with respect to $H_t$ (i.e., $L_{d^\prime}=L_{d}^\dag e^{-\frac{\beta}{2} \Delta_{d}}$ with $d^\prime$ being uniquely determined from $\Delta_{d} =-\Delta_{d^{\prime}}$). 
The effect of continuous measurement is represented by the Lindblad operators $\{M_y\}$, and the feedback is performed by changing the Hamiltonian.
In the following, we give a more detailed explanation on continuous measurement and feedback.

Continuous measurement is the readout of system's information via interaction with the measurement apparatus (e.g., the monitoring of an emitted photon from the system).
The measurement outcome obtained at $t_{n+1}$ is denoted as $y_{n+1}$ with the corresponding quantum jump represented by $M_{y_{n+1}}$ (e.g., the detection of a photon). If any measurement jump does not occur at $t_{n+1}$, $y_{n+1}$ is defined as $0$. 
The conditional dynamics of the system is described by Eq.~\eqref{SME}, where the Poisson increment $\Delta N_{y}$ is defined as $\Delta N_{y} =1$ if the jump $M_y$ occurs, and $\Delta N_{y} =0$ otherwise \cite{wiseman2009quantum}. 
We here emphasize that, by taking the ensemble average over the outcomes $Y_{n+1}$, Eq.~\eqref{SME} reduces to the ordinary master equation that describes dynamics interacting with an external system without post-selection of the measurement results. Such a decomposition that allows us to recover the master equation is called \emph{unraveling}.

Continuous feedback is provided by varying $H_t$ and/or $h_t$ according to the measurement results.
Because of the causality, the Hamiltonians in $[t_n,t_{n+1})$ are completely determined by measurement results before $t_n$ (i.e., $Y_n$) while it does not depend on those after $t_n$.
In our setup, the following types of the Hamiltonian variations are supposed: adiabatic change of the system Hamiltonian $H_t$ \cite{albash2012quantum,bulnes2016quantum}, perturbation of the external field (i.e., $h_t \ll H_t$) \cite{carmichael1999statistical,silaev2014lindblad,liu2014equivalence}, and sequential pulses (i.e., $h_t =\sum_i v_i \delta (t-s_i)$).
Hamiltonian variations other than those may not be given in the form of Eq.~\eqref{SME} \cite{szczygielski2013markovian,cuetara2015stochastic}, and hence are excluded in the following argument.
Note that the dependence on \begin{math}Y_n\end{math} of some operators and variables (such as \begin{math}H_t,h_t,L_{d},\Delta_{d},\Delta N_{y}\end{math}) is abbreviated for simplicity. 

\emph{Generalized second law.---}
In this setup, the ensemble average of thermodynamic quantities such as the heat $\langle Q\rangle$ dissipated to the heat bath and the entropy change $\langle \Delta S\rangle$ between the initial and final states can be defined as follows \cite{schaller2014open,ptaszynski2019thermodynamics,gong2016quantum}:
\begin{equation}
    \label{heat_entprd}
    \begin{split}
         \langle Q\rangle &\equiv \sum_{n=0}^{N-1}\sum_{Y_{n}}P[Y_{n}]\sum_{d}\tr[H_{t_n}\mathcal{D}[L_{d}]\rho_{t_n}^{Y_{n}}]\Delta t,\\
    \langle\Delta S\rangle &\equiv S(\rho_{\tau}) -S(\rho_0), 
    \end{split}
\end{equation}
where $\rho_0$ and $\rho_\tau \equiv \sum_{Y_N}P[Y_N]\rho_\tau^{Y_N}$ are the initial and final density operators, and $S(\rho) \equiv -\tr[\rho\ln \rho]$ represents the von Neumann entropy.
The average entropy production is defined as $\langle \sigma \rangle \equiv \langle\Delta S\rangle +\beta \langle Q\rangle$.

We now introduce the QC-transfer entropy as
\begin{equation}
    \label{QC_transfer_entropy}
    \langle i_{\mathrm{QC}} \rangle = \sum_{n=0}^{N-1}  \sum_{Y_{n}} P[Y_{n}]\mathcal{I}_{\mathrm{QC}}(\rho_{t_n}^{Y_{n}}:y_{n+1}),
\end{equation}
where $\mathcal{I}_{\mathrm{QC}}$ represents the QC-mutual information. Here, $\mathcal{I}_{\mathrm{QC}}$ is defined as $\mathcal{I}_{\mathrm{QC}}(\rho:y) \equiv S(\rho) -\sum_y P[y] S(\rho^y)$, where $P[y]$ is the probability of the outcome $y$, and $\rho^y$ denotes the conditional density operator after the measurement of $y$.
QC-mutual information quantifies the information obtained by measurement \cite{groenewold1971problem,ozawa1986information} so that it gives the upper bound of the accessible classical information by quantum measurement \cite{buscemi2008global,sagawa2013second}, and also has an operational interpretation through an informational task called measurement compression \cite{winter2004extrinsic,Wilde2012,Berta2014}.
On the basis of the foregoing definitions, the SL is generalized as
\begin{equation}
    \label{sgeneralized SL} \langle \sigma\rangle \geq -\langle i_{\mathrm{QC}}\rangle.
\end{equation}
This inequality gives the lower bound of $\langle \sigma\rangle$ under continuous measurement and feedback and reveals the relationship between the entropy production and quantum information at the level of ensemble average.

We here discuss the relationship between the QC-transfer entropy and the (classical) transfer entropy \cite{schreiber2000measuring} (Fig.~\ref{fig:setup}(a)). The transfer entropy is defined as
\begin{equation}
\label{Transfer entropy}
     \langle i_{\rm TE} \rangle \equiv \sum_{n=0}^{N-1}I(x_{n}:y_{n+1}|Y_{n}), 
\end{equation}
where $x_n$ denotes a state of a classical system at $t_n$. We define $I$ as the conditional mutual information $I(x_{n}:y_{n+1}|Y_{n})\equiv \sum_{Y_n}P[Y_n](\mathcal{H}_{Y_n}(x_n)-\sum_{y_{n+1}}P[y_{n+1}|Y_n] \mathcal{H}_{Y_n,y_{n+1}}(x_n))$, where $\mathcal{H}_{Y_n}(x_n)$ denotes the Shannon entropy of $x_n$ when the measurement results are $Y_n$.
From Eqs.~\eqref{QC_transfer_entropy} and \eqref{Transfer entropy}, we can see that in classical systems, information transfer in $[t_n,t_{n+1})$ is described by the conditional mutual information $I(x_{n}:y_{n+1}|Y_{n})$, whereas in quantum systems it is represented by the QC-mutual information of the conditional density operator $\rho_{t_n}^{Y_n}$.
Therefore, $\langle i_{\rm QC}\rangle$ is the quantum counterpart of $\langle i_{\rm TE} \rangle$, in that they both represent the total information gain obtained by accumulating conditional information transfer over time. 

\emph{Generalized fluctuation theorem.---}
We next introduce the generalized FT under continuous measurement and feedback, which is the main result of this Letter. Since the generalized FT is the equality concerning the stochastic entropy production $\sigma$ and the stochastic information gain $i_{\rm QC}$, we need to introduce proper definitions of these quantities. 
On the basis of a special class of stochastic decomposition of Eq.~\eqref{SME}, which we name \emph{fine unraveling}, both of these quantities can be defined for individual unraveled trajectories, which we call fine trajectories. 
In the following, we elaborate on these concepts.

\begin{figure}[tb]
\begin{center}
\includegraphics[width=0.45\textwidth]{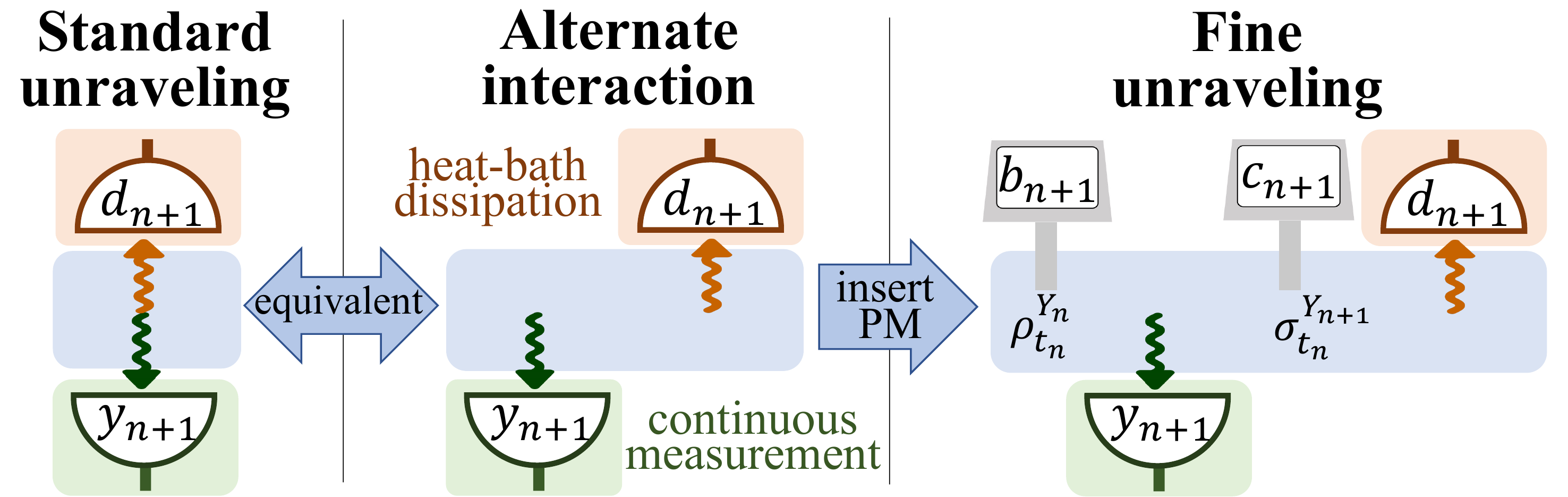}
\caption{The illustration of the standard unraveling, the alternate interaction situation, and the fine unraveling in $[t_{n},t_{n+1})$. The stochastic dynamics of the standard unraveling and the alternate interaction situation coincide in the continuous time limit $\Delta t\to 0$. The fine unraveling is introduced by inserting projective measurements $b_{n+1}$ and $c_{n+1}$ between the alternate interaction with the measurement apparatus and the heat bath.}
\label{fig:unravel}
\end{center}
\end{figure}

As the preliminary step toward defining the fine unraveling, we first introduce another unraveling, which we call \emph{standard unraveling}.
In this unraveling, we monitor the heat-bath dissipation in addition to the original continuous measurement of $y_{n+1}$ in Eq.~\eqref{SME}. We here define $\Delta N_d$ as the Poisson increment of $L_d$ and $d_{n+1}$ as the outcome of the heat-bath monitoring in $[t_n,t_{n+1})$.
We further perform two-time projective measurement at $t=0$ and $\tau$, just as in the standard scheme in stochastic thermodynamics \cite{funo2018quantum,sagawa2013second}.
Here, the two-time measurement is performed in the diagonalized bases of the density operators $\rho_0\equiv \sum_{a_0}p_0(a_0)\ket{a_0}\bra{a_0}$ and $\rho_\tau \equiv \sum_{a_\tau}p_\tau(a_\tau)\ket{a_\tau}\bra{a_\tau}$, and their outcomes are denoted as $a_0$ and $a_\tau$, respectively.
Thus, the dynamics of Eq.~\eqref{SME} is decomposed according to the measurement outcomes $\psi_\tau \equiv (a_0,a_\tau,\{y_n\}_{n=1}^{N},\{d_n\}_{n=1}^{N})$.
We refer to the unraveled trajectory designated by $\psi_\tau$ as the standard trajectory.
We remark that Eq.~\eqref{SME} can be recovered from the standard unraveling by taking the ensemble average over the results of the two-time measurement $a_0,a_\tau$ and the heat-bath monitoring $\{d_n\}$.

The fine unraveling is introduced by the following two-step transformation from the standard unraveling (see Fig.~\ref{fig:unravel}).
We first consider the situation that a measurement and an interaction with the heat bath occur alternately every $\Delta t$, and then insert the projective measurements in the diagonalized bases of \begin{math}\rho_{t_n}^{Y_n}\end{math} and \begin{math}\sigma_{t_n}^{Y_{n+1}}\end{math} right before and after the measurement of $y_{n+1}$, respectively. The diagonalizations are defined as $\rho_{t_n}^{Y_n} \equiv \sum_{b_{n+1}}p^{Y_n}(b_{n+1})\ket{b_{n+1}}\bra{b_{n+1}}$ and $\sigma_{t_n}^{Y_{n+1}} \equiv \sum_{c_{n+1}}p^{Y_{n+1}}(c_{n+1})\ket{c_{n+1}}\bra{c_{n+1}}$, and the outcomes of the inserted measurements before and after the monitoring of $y_{n+1}$ are $b_{n+1}$ and $c_{n+1}$, respectively.
Here, \begin{math}\sigma_{t_n}^{Y_{n+1}}\end{math} represents the conditional density operator when the measurement results are $Y_{n+1}$ in the alternate interaction situation.  
Thus, the fine unraveling decomposes Eq.~\eqref{SME} according to $\psi_\tau$ and $\pi_\tau$, where $\pi_\tau \equiv (\{b_n\}_{n=1}^{N},\{c_n\}_{n=1}^{N})$ denotes the outcomes of the inserted projective measurements altogether.
By taking the ensemble average over $\pi_\tau$, along with $a_0,a_\tau$ and $\{d_n\}$, the fine unraveling reproduces the original dynamics of Eq.~\eqref{SME}. It should be emphasized that these projective measurements do not destroy the measured states at the ensemble average level, because of their choices of the bases.

For each fine trajectory, we can define both the stochastic entropy production and the QC-transfer entropy, and derive the generalized FT.
The stochastic entropy production is defined as $\sigma[\psi_\tau,\pi_\tau] \equiv\Delta S [\psi_\tau,\pi_\tau]+\beta Q[\psi_\tau,\pi_\tau]$, where the stochastic heat and stochastic entropy change are defined as follows \cite{gong2016quantum,horowitz2012quantum,horowitz2013entropy,hekking2013quantum,liu2016calculating,manzano2021quantum}:
\begin{equation}
\label{letter:QdStraj}
\begin{split}
     Q[\psi_\tau,\pi_\tau] &\equiv \sum_{n=0}^{N-1}\sum_{d} \Delta N_{d} \Delta_{d},\\
     \Delta S[\psi_\tau,\pi_\tau] &\equiv -\ln{p_\tau(a_\tau)}+\ln{p_0(a_0)}.
\end{split}
\end{equation}
The stochastic QC-transfer entropy is then defined as 
\begin{equation}
\label{stochatic QC transfer entropy}
    i_{\mathrm{QC}}[\psi_\tau,\pi_\tau] \equiv \sum_{n=0}^{N-1} -\ln p^{Y_n}(b_{n+1}) +\ln p^{Y_{n+1}}(c_{n+1}).
\end{equation}
We can confirm that the ensemble averages of the stochastic quantities \eqref{letter:QdStraj} and \eqref{stochatic QC transfer entropy} coincide with Eqs.~\eqref{heat_entprd} and \eqref{QC_transfer_entropy}.
Based on the foregoing definitions, the generalized FT is expressed as 
\begin{equation}
    \label{generalized_fluctuation_theorem}
    \langle e^{-\sigma -i_{\rm QC}} \rangle =1.
\end{equation}
This equality implies the generalized SL \eqref{sgeneralized SL} and reveals the relationship between the entropy production and the QC-transfer entropy at the trajectory level. See Supplemental Material for the full proof of Eq.~\eqref{generalized_fluctuation_theorem}~\footnote{See Supplemental Material for details}.

\emph{Protocol of the hybrid method.---}
\begin{figure}[tb]
\begin{center}
\includegraphics[width=0.45\textwidth]{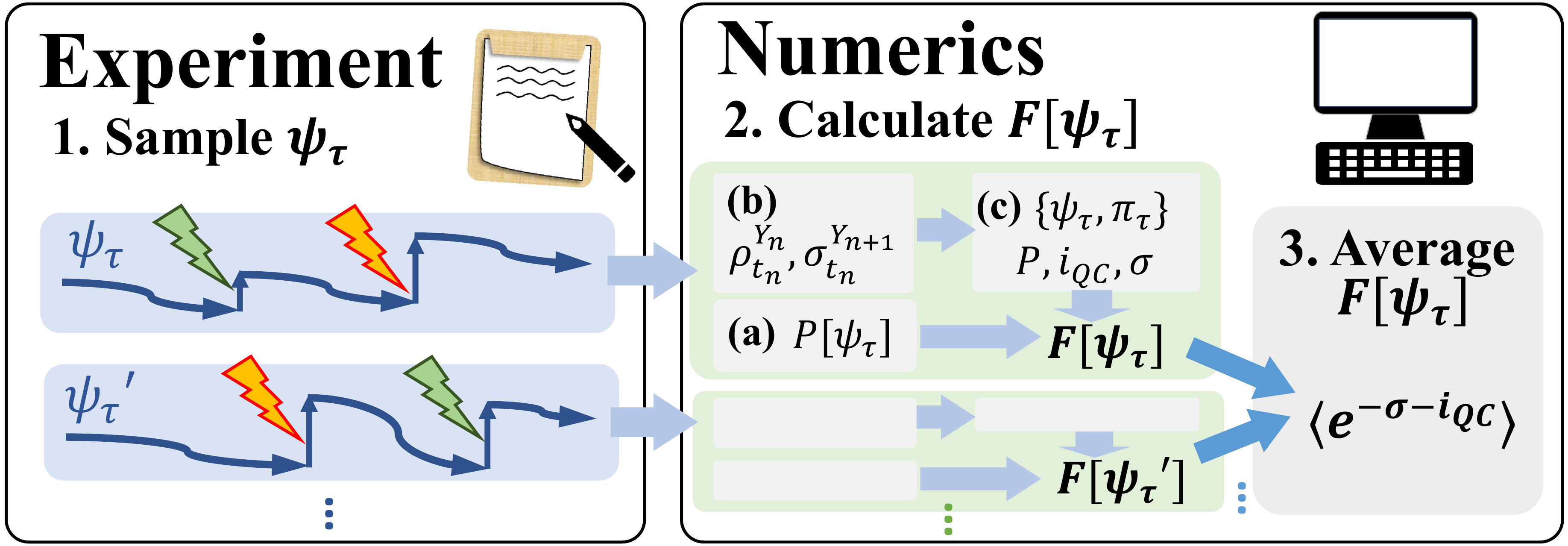}
\caption{The schematics for the experiment-numerics hybrid verification method. The verification protocol is composed of the experimental part, in which the standard trajectories $\psi_\tau$ are sampled, and the numerical calculation part, in which $F[\psi_\tau]$ are calculated, and then their average value is taken. The average value coincides with the left-hand side of Eq.~\eqref{generalized_fluctuation_theorem}.}
\label{fig:expnum_letter}
\end{center}
\end{figure}
Although the direct realization of the fine unraveling in real experiments is difficult, the generalized FT (Eq.~\eqref{generalized_fluctuation_theorem}) can be verified by an experiment-numerics hybrid verification method, in which the standard trajectories are sampled in an experiment and the auxiliary numerical calculation is performed in a classical computer (Fig.~\ref{fig:expnum_letter}).
The detection of the standard trajectories is feasible in real experiments such as circuit QED \cite{gasparinetti2015fast,viisanen2015incomplete,karimi2020quantum,karimi2020reaching} and cavity QED \cite{guerlin2007progressive,deleglise2008reconstruction}.
Then, by the auxiliary numerical calculation, we can evaluate the left-hand side of Eq.~\eqref{generalized_fluctuation_theorem}.
The concrete protocol of the hybrid method is as follows:
\begin{enumerate}
    \item By a real quantum experiment, sample the standard trajectories.
    \item By classical numerical simulation, calculate \begin{math}F[\psi_\tau] \equiv \sum_{\pi_\tau}\frac{P[\psi_\tau,\pi_\tau]}{P[\psi_\tau]}e^{-\sigma[\psi_\tau,\pi_\tau]-i_{\mathrm{QC}}[\psi_\tau,\pi_\tau]}\end{math} for each experimentally sampled trajectory $\psi_\tau$:
    \begin{enumerate}
        \item Calculate the realization probability $P[\psi_\tau]$ by solving the stochastic Schr\"odinger equation of the standard unraveling.
        \item Calculate the dynamics of conditional density operators $\rho_{t_n}^{Y_n}$ and $\sigma_{t_n}^{Y_{n+1}}$ by solving Eq.~\eqref{SME}.
        \item Calculate $P[\psi_\tau,\pi_\tau]$, $i_{\mathrm{QC}}[\psi_\tau,\pi_\tau]$ and $\sigma[\psi_\tau,\pi_\tau]$ for all the corresponding fine trajectories $\{\psi_\tau,\pi_\tau\}_{\pi_\tau}$ by using the solution of (b).
    \end{enumerate}
    \item Average $F[\psi_\tau]$ over all sampled trajectories, which gives the left-hand side of Eq.~\eqref{generalized_fluctuation_theorem}.
\end{enumerate}

We make some remarks on the second step above. 
While $P[\psi_\tau]$ can be calculated solely from the standard unraveled dynamics, we have to prepare the inserted projective measurements in order to calculate the quantities for fine trajectories $P[\psi_\tau,\pi_\tau]$, $i_{\mathrm{QC}}[\psi_\tau,\pi_\tau]$ and $\sigma[\psi_\tau,\pi_\tau]$. 
The number of realizable fine trajectories with $P[\psi_\tau,\pi_\tau]\neq0$ calculated in (c) is finite even in the limit of $\Delta t\to 0$, because the outcomes of inserted projective measurements in the fine unraveling changes only right after any quantum jump $\{y_n\}$ or $\{d_n\}$ occurs \cite{Note1}. Therefore, the exact calculation of $F[\psi_\tau]$ can be performed with reasonable numerical cost.

\emph{Numerical demonstration of the generalized FT.---} 
\begin{figure}[tb]
\begin{center}
\includegraphics[width=0.45\textwidth]{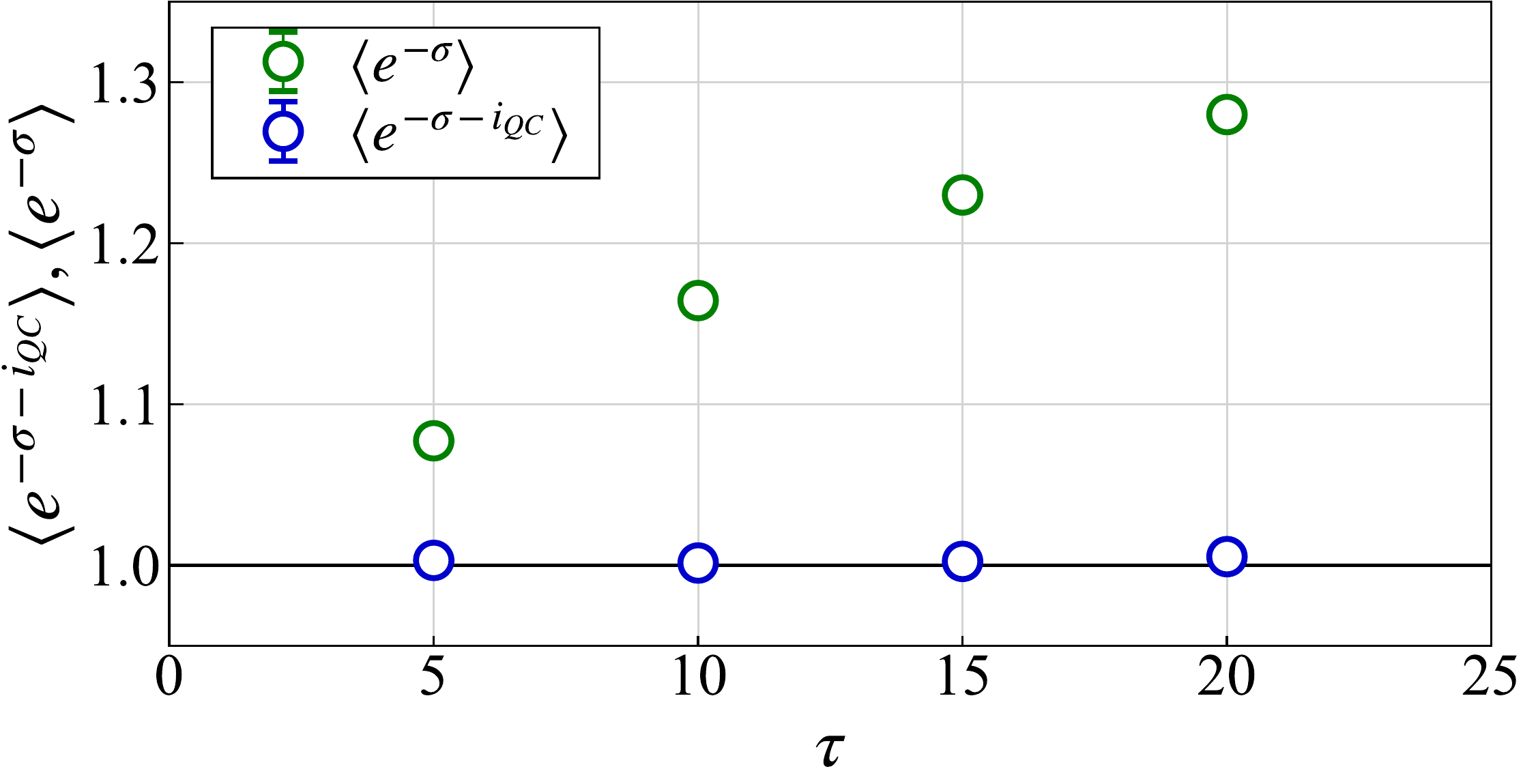}
\caption{Numerical verification of the generalized FT. The average values of \begin{math}e^{-\sigma}\end{math} and \begin{math}e^{-\sigma-i_{\mathrm{QC}}}\end{math} for the sampled fine trajectories are plotted.  
The system parameters are taken as \begin{math}\omega = 0.3,\epsilon = 0.04, \omega_0 = 0.1\pi, \beta = 1, \gamma_{\pm} = 0.015\omega \{\coth{(\frac{\beta \omega}{2})}\mp 1\}, \gamma_m = \gamma_{+}+\gamma_{-}\end{math} and $\delta =0.2$. Each data point denotes the average over \begin{math}1.0 \times 10^{5}\end{math} trajectories.}
\label{fig:fluctuation}
\end{center}
\end{figure}
To further support our findings, we have numerically calculated the fine unraveled dynamics of the two-level system to verify the generalized FT. We employ the setting that the population of excited state $\ket{1}$ is reduced by continuous measurement and feedback. The system Hamiltonian is fixed as \begin{math}H_t=\omega \sigma_z\end{math}, the coherent driving is applied as $h_t=\epsilon \sigma_x \cos \omega_0 t$, the heat-bath dissipation is represented by \begin{math}L_{\pm} = \sqrt{\gamma_{\pm}} \sigma_{\pm} \end{math}, and the continuous measurement operator is defined as \begin{math}M_1 = \sqrt{\gamma_m} (\ket{1}\bra{1}+ \delta \ket{0}\bra{0} + \delta \sigma_x) \end{math}, where $\sigma_i$ denotes the Pauli matrix. 
We note that if $\delta$ is negligible ($\delta \ll 1$), we can almost certainly decide that the system is in $\ket{1}$ after the detection of $M_1$. Thus we apply unitary gate $\sigma_x$ right after the detection, in order to reduce the excited state population. If the detection occurs at $t_n$, the feedback protocol is realized by applying the pulse in $t_{n+1}$, which changes the external driving Hamiltonian as $h_t=\epsilon \sigma_x\cos \omega_0 t +v\delta(t-t_{n+1})$ with $e^{-iv}\equiv\sigma_x$. 
By computing $\sigma$ and $i_{\mathrm{QC}}$ for individual trajectories, we have verified that the generalized FT holds.
This is illustrated in Fig.~\ref{fig:fluctuation}, where we can see that \begin{math}\langle e^{-\sigma}\rangle\end{math} increases with \begin{math}\tau\end{math}, implying the violation of the conventional FT (i.e., \begin{math}\langle e^{-\sigma}\rangle \neq 1\end{math}). 
We also provide numerical results regarding the time evolution of $i_{\rm QC}$ and the verification of the generalized SL in Supplemental Material \cite{Note1}.

\emph{Summary and outlook.---}
In this Letter, we have addressed a general principle of nonequilibrium thermodynamics in quantum systems under continuous measurement and feedback. The generalized SL \eqref{sgeneralized SL} and FT \eqref{generalized_fluctuation_theorem} reveal the relationship between the entropy production and the quantum-information gain at the ensemble level and trajectory level, respectively. The newly introduced information gain, the QC-transfer entropy, would play an important role in quantifying quantum information transfer by sequential quantum measurements.
We have also verified the generalized FT by numerical simulation and proposed a feasible experiment-numerics hybrid verification method.

We here make several remarks related to the results in this Letter, whose detailed explanations are given in Supplemental Material \cite{Note1}.
First, when absolute irreversibility \cite{funo2015quantum,murashita2017fluctuation} is caused by continuous measurement, the correction term $\lambda_{\rm irr}>0$ must be incorporated to the generalized FT as $\langle e^{-\sigma-i_{\mathrm{QC}}}\rangle =1-\lambda_{\rm irr}$. We can derive a simple sufficient condition for $\lambda_{\rm irr} =0$ in our setup.
Second, we can also derive the generalized FT under continuous measurement with imperfect detection rate $\eta_{y}<1$ in the same way as Eq.~\eqref{generalized_fluctuation_theorem}. The numerical simulation in such setup demonstrates that $\langle e^{-\sigma}\rangle$ decreases with the detection rate, while the generalized FT \eqref{generalized_fluctuation_theorem} always holds.
Finally, the generalized FT can also be derived under the standard unraveling as $\langle e^{-\sigma-i_{\rm QJT}} \rangle =1$, by introducing another information term $i_{\rm QJT}$ \cite{gong2016quantum}.

We show some future perspectives related to our work.
First, it is important to verify the generalized FT (Eq.~\eqref{generalized_fluctuation_theorem}) in real experimental systems by using the experiment-numerics hybrid verification method.
Another interesting future task is to generalize the FT under the other type of continuous measurement described by Wiener process \cite{wiseman2009quantum}.
Finally, it is also intriguing to clarify the difference between quantum-information flow that occurs spontaneously in many-body quantum systems \cite{ptaszynski2019thermodynamics} and the QC-transfer entropy under explicit measurement and feedback as in the present work.

\emph{Acknowledgements.---}T.Y. is supported by World-leading Innovative Graduate Study Program for Materials Research, Industry, and Technology (MERIT-WINGS) of the University of Tokyo. 
N.Y. wishes to thank JST PRESTO No. JPMJPR2119.
T.S. is supported by JSPS KAKENHI Grant Number JP19H05796 and JST, CREST Grant Number JPMJCR20C1, Japan. T.S. is also supported by Institute of AI and Beyond of the University of Tokyo.

\bibliography{bibliography.bib}

\onecolumngrid

\clearpage
\begin{center}
	\Large
	\textbf{Supplemental Material for ``Quantum Fluctuation Theorem under Continuous Measurement and Feedback"}
\end{center}

\setcounter{section}{0}
\setcounter{equation}{0}
\setcounter{figure}{0}
\setcounter{table}{0}
\renewcommand{\thesection}{S\arabic{section}}
\renewcommand{\theequation}{S\arabic{equation}}
\renewcommand{\thefigure}{S\arabic{figure}}
\renewcommand{\thetable}{S\arabic{table}}

\section{Our setup}
In this section, we introduce our setup and the definitions of the thermodynamic quantities at the ensemble level. In Section \ref{dynamics at the ensemble level}, we briefly explain our setup and give some remarks on the derivation of the stochastic master equation (SME). In Section \ref{Thermodynamic quantities at the ensemble level}, we show the definitions of the thermodynamic quantities in our setup. We note that we basically consider the system under continuous measurement with imperfect detection rate (i.e., $\eta_y \leq 1$) in Supplemental Material. This includes the perfect detection rate (i.e., $\eta_y = 1$), which is discussed in the main text, as the special case.

\subsection{Dynamics at the ensemble level}\label{dynamics at the ensemble level}
In this work, we consider a quantum system interacting with the heat bath at inverse temperature $\beta$ under continuous measurement with arbitrary detection rate $\{\eta_{y}\}$. We descritize time as $t_n\equiv n\Delta t$ and consider the time evolution from $t=0$ to $t=\tau \equiv t_N$, and then take the continuous time limit $\Delta t \to 0, N\to \infty$ while keeping $\tau$ constant. As explained in the main text, the dynamics is basically represented by the SME \eqref{SME}, but occasionally unitary gates caused by sudden pulses are inserted. 
The application of the sudden pulse $h_t=v_i\delta (t-s_i)$ means that the system follows the SME in $[0,s_i)$ and $[s_i,\tau)$, while the density operator is changed abruptly at $t=s_i$ as $\rho\to V_i \rho V_i^\dag $, where $ V_i$ is defined as $ V_i \equiv e^{-iv_i}$.
Note that the notation $h_t=v_i\delta (t-s_i)$ is just used for simplicity and does not faithfully mean the stochastic differential equation with discontinuously changed coefficient, which does not have a unique solution.
We assume that the pulse application time $s_i$ can be described as $s_i=n_i\Delta t$. In other words, the dynamics can be described just by inserting the unitary gate in between the time intervals $[t_{n_i-1},t_{n_i})$ and $[t_{n_i},t_{n_i+1})$.

In this setup, the dynamics is described by the following SME and unitary gate insertion:
\begin{equation}
    \label{SME_eta}
    \begin{split}
       \rho_{t_{n+1}}^{\prime Y_{n+1}} &= \rho_{t_{n}}^{Y_{n}}+ \Delta t\Bigl\{-i[H_{t_n}^{Y_n}+h_{t_n}^{Y_n},\rho_{t_n}^{Y_n}] + \sum_{d}\mathcal{D}[L_{d}^{Y_n}]\rho_{t_n}^{Y_n} +\sum_{y} (1-\eta_{y}) \mathcal{D} [M_{y}]\rho_{t_n}^{Y_n} \\
    &+ \sum_{y}-\frac{\eta_{y}}{2}\{M_{y}^\dag M_{y},\rho_{t_n}^{Y_n}\} + \eta_{y}\tr[M_{y}\rho_{t_n}^{Y_n}M_{y}^\dag]\rho_{t_n}^{Y_n}\Bigr\} +\sum_{y} \Delta N_{y,\rm det}^{Y_n} \mathcal{G} [M_{y}] \rho_{t_n}^{Y_n},\\
    \end{split}
\end{equation}
where \begin{math}\mathcal{D}[c]\rho\equiv c\rho c^\dag -1/2\{c^\dag c,\rho\}\end{math} and \begin{math}\mathcal{G}[m]\rho \equiv (m\rho m^\dag /\tr[ m\rho m^\dag]) -\rho\end{math}, and
\begin{equation}
    \label{pulse_unitary}
    \begin{split}
     \rho_{t_{n+1}}^{Y_{n+1}} &\equiv V^{Y_n}\rho_{t_{n+1}}^{\prime Y_{n+1}}V^{Y_n\dag},\\
    V^{Y_n} &\equiv
    \begin{cases}
    1 &(\mathrm{no\  pulse\  application\ at}\ t_{n+1}) \\
    e^{-iv_{t_{n+1}}} &(\mathrm{sudden\  pulse\ } v_{t_{n+1}}\delta (t-t_{n+1})\  \mathrm{applied\  at\ } t_{n+1}).
    \end{cases}
    \end{split}
\end{equation}
Here, \begin{math}y_n\end{math} is the outcome obtained from continuous measurement in \begin{math}[t_{n-1},t_{n})\end{math}, \begin{math}Y_n \equiv (y_1,y_{2},\dots ,y_{n})\end{math} is all the outcomes obtained from the  measurement before \begin{math}t_n\end{math}, and \begin{math}\rho_{t_n}^{Y_n}\end{math} is the density operator at $t_n$ when the measurement results are \begin{math}Y_n\end{math}. We define $H_{t_n}^{Y_n}$ as the intrinsic system Hamiltonian, $h_{t_n}^{Y_n}$ as the perturbative external driving Hamiltonian, $V^{Y_n}$ as the unitary gate caused by the sudden pulse, and $\rho_{t_{n+1}}^{\prime Y_{n+1}}$ as the conditional density operator before the pulse application at $t_{n+1}$. Continuous feedback is performed by changing $H_{t_n}^{Y_n}$ and $h_{t_n}^{Y_n}$, and applying the sudden pulse $V^{Y_n}$ according to the measurement results $Y_n$.
The Lindblad operators $\{L_{d}^{Y_n}\}$ represent the dissipation due to the interaction with the heat bath, and $\{M_{y}\}$ represent continuous measurement with the detection rate $\{\eta_{y}\}$. The Poisson increment \begin{math}\Delta N_{y,\rm det}^{Y_n}\end{math} is 1 (and the measurement result $y_{n+1} =y$ is obtained) in the probability of \begin{math}\eta_{y}\tr[M_{y}\rho_{t_n}^{Y_n}M_{y}^{\dag}]\Delta t\end{math}, and \begin{math}\Delta N_{y,\rm det}^{Y_n}=0\end{math} otherwise. The SME provided in the main text (Eq.~\eqref{SME}) is a special case (i.e., \begin{math}\eta_{y}=1\end{math}) of SME~\eqref{SME_eta}. In the following, we will give the detailed explanation of the measurement outcome $y_n$ and the heat-bath dissipation $\{L_d^{Y_n}\}$. After that, we will make some remarks on the assumptions necessary to describe the system's dynamics by Eq.~\eqref{SME_eta} and \eqref{pulse_unitary}.

Strictly speaking, we define $y_n$ as the outcome of continuous measurement which is read out via the interaction in $[t_{n-1},t_n)$.
If we take into account the delay time of the measurement $\tau_{\rm delay}$, $y_n$ is obtained at $t_n+\tau_{\rm delay}$. The delay time corresponds, for example, to the time for a photon to travel from the system to the photodetector. Therefore, only the measurement results until $t_{n-n'}$ (i.e., $Y_{n-n'}$) can be fed back to the Hamiltonian in $[t_n,t_{n+1})$ in such cases, where $n'$ is defined as $n' \Delta t \leq \tau_{\rm delay} < (n'+1)\Delta t$. In the main text, we define $y_n$ as the measurement result newly obtained at $t_{n}$ under the assumption that the delay time is sufficiently smaller than the smallest time scale of the system dynamics (i.e., $\tau_{\rm delay} \ll \Delta t$). We note that our results remain valid even if the delay time is not negligible, since the assumption required for the proof is only that the Hamiltonian in $[t_n,t_{n+1})$ does not depend on the results after $t_n$ (i.e., $(y_{n+1},\dots ,y_{N})$).

The Lindblad operators of the heat-bath dissipation $\{L_d^{Y_n}\}$ are composed of the operators representing the transition between the energy eigenstates, and those representing the pure dephasing with no energy relation \cite{albash2012quantum}. The eigenstates transition operator is given by $L_d^{Y_n} \propto \ket{E_k^{Y_n}}\bra{E_l^{Y_n}}$, where $\ket{E_k^{Y_n}}$ denotes an eigenstate of $H_{t_n}^{Y_n}$ with energy $E_k^{Y_n}$, which satisfies $[L_d^{Y_n},H_{t_n}^{Y_n}] \equiv \Delta_{d}^{Y_n}L_d^{Y_n}$ and the detailed balance condition $L_{d^\prime}^{Y_n} =L_{d}^{Y_n\dag} e^{-\frac{\beta}{2}\Delta_{d}^{Y_n}}$, with $L_{d^\prime}^{Y_n} \propto \ket{E_l^{Y_n}}\bra{E_k^{Y_n}}$ being the inverse transition of $L_d^{Y_n}$.
The pure dephasing operator satisfies $[L_d^{Y_n},H_{t_n}^{Y_n}] =0$ and is Hermitian $L_d^{Y_n\dag}=L_d^{Y_n}$. Since no energy dissipation occurs in the pure dephasing (i.e., $\Delta_d^{Y_n} =0$), $L_d^{Y_n}$ satisfies the detailed balance condition with itself.

We here introduce two assumptions in order to represent the system's dynamics by Eq.~\eqref{SME_eta} and \eqref{pulse_unitary}.
First, we assume that the Lindblad generator can be additively decomposed to the contributions from the heat-bath dissipation and continuous measurement.
This assumption holds if the system-bath interaction and the interaction between the system and the measurement apparatus are both weak enough \cite{schaller2014open,bulnes2016quantum,ptaszynski2019thermodynamics}, while the additivity can be violated in the region beyond the weak coupling regime.
Secondly, we derive Eq.~\eqref{SME_eta} by applying the Born-Markov and rotating-wave approximations. These approximations are valid and the dynamics is described in the form of Eq.~\eqref{SME_eta} in the case that the variation of $H_{t_n}^{Y_n}$ is adiabatic and $h_{t_n}^{Y_n}$ is perturbative, since the system-bath interaction is almost unaffected by such time variations \cite{albash2012quantum}.
The application of sudden pulses $v_{t_{n+1}}\delta (t-t_{n+1})$ does not also affect the form of the SME \eqref{SME_eta} and can be represented just by inserting the unitary gates $V^{Y_n}\equiv e^{-iv_{t_{n+1}}}$ at $t=t_{n+1}$, if the timescale of the pulse application is so small that the time evolution by $H_{t_n}^{Y_n}$ and the dissipation and measurement quantum jumps $\{L_d^{Y_n}\},\{M_y^{Y_n}\}$ during the pulse application are negligible.
We note that the allowed variations of Hamiltonian $H^{Y_n}_{t_n}+h^{Y_n}_{t_n}$ in our work are the same as those of Ref.~\cite{gong2016quantum}.
\subsection{Thermodynamic quantities at the ensemble level}\label{Thermodynamic quantities at the ensemble level}
Under the SME \eqref{SME_eta}, the heat dissipated to the heat bath can be defined as 
\begin{equation}
    \label{Q_ens}
\langle Q\rangle \equiv \sum_{n=0}^{N-1} \big\{ \sum_{Y_n}P[Y_n]\sum_{d}\tr[L_{d}^{Y_n}\rho_{t_n}^{Y_n}L_{d}^{Y_n\dag}]\Delta t\Delta_{d}^{Y_n} \big\}.
\end{equation}
We here introduce \begin{math}P[Y_n]\end{math} as the probability that takes the measurement result \begin{math}Y_n\end{math} at the ensemble level dynamics. We note that $\langle Q\rangle$ only includes the energy dissipated to the heat bath, where the energy change caused by system Hamiltonian variation, external driving, and continuous measurement are excluded. 

The entropy change from $t=0$ to $\tau$ is defined as 
\begin{equation}
\label{dS_ens}
\langle \Delta S\rangle \equiv S(\rho_\tau) -S(\rho_0),
\end{equation}
where $\rho_0$ is the initial density operator and \begin{math}\rho_\tau \equiv \sum_{Y_N}P[Y_N]\rho_\tau^{Y_N}\end{math} is the final density operator. The entropy production is defined as $\langle \sigma \rangle \equiv \langle \Delta S\rangle +\beta \langle Q\rangle$.

\section{Standard unraveling and fine unraveling}\label{unraveling} 
In this section, we introduce two ways of unraveling, namely the standard unraveling and fine unraveling, and define thermodynamic quantities to each unraveled trajectory. Since the fine unraveling is newly introduced in this work, we give a detailed explanation on it. We confirm that the original SME \eqref{SME_eta} is reproduced by taking the ensemble average over the fine trajectories, and that the thermodynamic quantities defined to each fine trajectory are consistent with their definitions at the ensemble level (Eq.\eqref{Q_ens}, \eqref{dS_ens}). On the other hand, the detailed explanation on the standard unraveling, which is widely used in quantum thermodynamics, is abbreviated (refer to e.g., Ref.~\cite{horowitz2012quantum,horowitz2013entropy,gong2016quantum,hekking2013quantum,liu2016calculating,liu2014equivalence,manzano2021quantum}). The explanation on the standard unraveling is in Section \ref{Standard unraveling} and that of the fine unraveling is in Section \ref{fine unraveling}.
\subsection{Standard unraveling}\label{Standard unraveling}
The standard unraveling is realized by monitoring all the quantum jumps and performing the two-time projective measurement (TPM) at $t=0$ and $\tau$. We first explain the continuous monitoring of the quantum jumps. Second, we elaborate on the TPM scheme which is commonly used in stochastic thermodynamics. Finally we define the stochastic thermodynamic quantities for the standard trajectories.

When we continuously monitor all the Lindblad operators in the SME \eqref{SME_eta}, namely \begin{math}\{L_{d}^{Y_n}\}\end{math}, \begin{math}\{\sqrt{1-\eta_{y}}M_{y}\}\end{math}, and \begin{math}\{\sqrt{\eta_{y}}M_{y}\}\end{math}, the time evolution in $[t_n,t_{n+1})$ can be described by the stochastic Schr\"odinger equation (SSE) as
\begin{equation}
\label{SSE_eta}
\begin{split}
    \ket{\psi_{t_{n+1}}}^\prime =\Biggl[ 1- \Delta t &\Bigl\{i(H_{t_n}^{Y_n}+h_{t_n}^{Y_n}) +\frac{1}{2}\Bigl(\sum_{y} M_{y}^\dag M_{y}-\| M_{y}\ket{\psi_{t_n}}\|^2  +\sum_{d} L_{d}^{Y_n\dag} L_{d}^{Y_n} -\| L_{d}^{Y_n}\ket{\psi_{t_n}}\|^2 \Bigr)\Bigr\}\Biggr]\ket{\psi_{t_n}}  \\
    +\sum_{y} \Delta N_{y,\rm det}^{Y_n}&\frac{M_{y}\ket{\psi_{t_n}}}{\| M_{y}\ket{\psi_{t_n}}\|}+\sum_{y} \Delta N_{y,\rm mis}^{Y_n}\frac{M_{y}\ket{\psi_{t_n}}}{\| M_{y}\ket{\psi_{t_n}}\|}+ \sum_{d}  \Delta N_{d}^{Y_n} \frac{L_{d}^{Y_n}\ket{\psi_{t_n}}}{\| L_{d}^{Y_n}\ket{\psi_{t_n}}\|},
    \end{split}
\end{equation}
where the Poisson increments are given as $\mathbb{E}[\Delta N_{y,\rm det}^{Y_n}]=\eta_{y}\| M_{y}\ket{\psi_{t_n}}\|^2 \Delta t$, 
$\mathbb{E}[\Delta N_{y,\rm mis}^{Y_n}]=(1-\eta_{y})\| M_{y}\ket{\psi_{t_n}}\|^2 \Delta t$ and $\mathbb{E}[\Delta N_{d}^{Y_n}]=\| L_{d}^{Y_n}\ket{\psi_{t_n}}\|^2 \Delta t$. We note that the notation $\ket{\psi_{t_{n+1}}}^\prime$ means the state before the pulse application (see Eq.~\eqref{ket_pulse_application}). The subscript `det' and `mis' denote the detection of $M_y$ and the misdetection of $M_y$, respectively.
We here introduce $d_{n}$ to denote the monitoring result of the heat bath in $[t_{n-1},t_{n})$ and \begin{math}D_{n} \equiv (d_1,d_{2},\dots,d_{n})\end{math} to represent all the outcomes about the heat-bath dissipation until $t_n$. If the dissipation jump $L_d$ is monitored at $t_n$, then $d_n$ is defined as $d_n=d$, and if there is no quantum jump, then $d_n$ is defined as $d_n =0$.
We also introduce $z_n$ to denote the actual quantum jump caused by continuous measurement. If the measurement quantum jump $y$ is detected in continuous measurement in $[t_{n-1},t_n)$, $y_n$ and $z_n$ are defined as $y_n=y,z_n=y$. On the other hand, if the measurement jump is not detected while the jump $M_{y}$ actually occurs in $[t_{n-1},t_{n})$, $y_n$ and $z_n$ are defined as $y_n=0,z_n=y$. In the case of perfect measurement, $y_n$ and $z_n$ always coincide. We also define \begin{math}Z_n \equiv(z_1,z_{2},\dots,z_{n})\end{math} to represent the actual measurement quantum jumps until $t_n$. In the same manner as the dynamics at the ensemble level, the conditional quantum state at $t_{n+1}$ after the sudden-pulse application can be defined as
\begin{equation}
\label{ket_pulse_application}
    \ket{\psi_{t_{n+1}}} \equiv  V^{Y_n}\ket{\psi_{t_{n+1}}}^\prime.
\end{equation}
The set of Kraus operators for the standard unraveling $\{\mathcal{L}_{d_{n+1},y_{n+1},z_{n+1}}^{Y_n}\}_{d_{n+1},y_{n+1},z_{n+1}}$ is defined as
\begin{equation}
        \label{Kraus_Sch_unravel}
    \mathcal{L}_{d_{n+1},y_{n+1},z_{n+1}}^{Y_n} \equiv
    \begin{cases}
    V^{Y_n} U_{\mathrm{nojump}}^{Y_n} &(d_{n+1}=0,y_{n+1}=0,z_{n+1}=0)\\
    V^{Y_n} L_{d_{n+1}}^{Y_n}\sqrt{\Delta t} &(d_{n+1}\neq0,y_{n+1}=0,z_{n+1}=0)\\
    V^{Y_n} \sqrt{\eta_{y_{n+1}}} M_{y_{n+1}}\sqrt{\Delta t} &(d_{n+1}=0,y_{n+1}\neq0,z_{n+1}=y_{n+1})\\
    V^{Y_n} \sqrt{1-\eta_{z_{n+1}}} M_{z_{n+1}}\sqrt{\Delta t} &(d_{n+1}=0,y_{n+1}=0,z_{n+1}\neq0),\\
    \end{cases}
\end{equation}
where $ U_{\mathrm{nojump}}^{Y_n}$ is defined as $ U_{\mathrm{nojump}}^{Y_n}\equiv 1- i(H_{t_n}^{Y_n}+h_{t_n}^{Y_n})\Delta t -\frac{1}{2}\{\sum_d L_d^{Y_n\dag}L_d^{Y_n} +\sum_{y} M_y^\dag M_y\}\Delta t$.

We further perform the TPM in order to define the entropy change \begin{math}\Delta S\end{math} for each trajectory. This is a standard scheme in quantum thermodynamics and widely used in the derivation of the fluctuation theorem (FT) and the generalized FT in previous works \cite{funo2018quantum,sagawa2013second}. In the TPM scheme, projective measurements are performed on the initial and final density operators. In particular, we perform the entropy-defining TPM, which is the projective measurements in the diagonal bases of \begin{math}\rho_0,\rho_\tau\end{math} at $t=0$ and $\tau$, respectively.
We define the diagonalizations of the initial and final density operators as \begin{math}\rho_0 \equiv \sum_{a}p_0(a)\ket{a}^{\!0}{}^{0\!\!}\bra{a}\end{math} and \begin{math}\rho_\tau \equiv \sum_{a}p_\tau(a)\ket{a}^{\!\tau}{}^{\tau\!\!}\bra{a}\end{math}, and define the measurement outcomes of the TPM at $t=0$ and $\tau$ as $a_0$ and $a_\tau$. We abbreviate the superscripts $0,\tau$ of the eigenstates in obvious cases. We here introduce \begin{math}\psi_{t_n}\equiv(a_0,D_n,Z_n,Y_n)\end{math} to represent all the outcomes which designate the single trajectory until $t=t_n$ (note that $\psi_\tau$ includes $a_\tau$).
The important point is that we can reconstruct the dynamics of SME \eqref{SME_eta} by taking the ensemble average of $\{d_n\}$, $\{z_n\}$ and $a_0,a_\tau$. This guarantees that the set of Kraus operators \eqref{Kraus_Sch_unravel} is an unraveling of the original dynamics at the ensemble level.

The stochastic thermodynamic quantities, heat $Q[\psi_\tau]$, entropy change $\Delta S [\psi_\tau]$ and entropy production $\sigma[\psi_\tau]$, are defined for each trajectory as
\begin{equation}
    \label{Q_Sch}
    Q[\psi_\tau] \equiv \sum_{n=0}^{N-1}\sum_{d_{n+1}} \Delta N_{d_{n+1}}^{Y_n} \Delta_{d_{n+1}}^{Y_n},
\end{equation}
\begin{equation}
    \label{dS_Sch}
    \Delta S[\psi_\tau] \equiv -\ln{p_\tau(a_\tau)}+\ln{p_0(a_0)},
\end{equation}
\begin{equation}
    \label{sig_Sch}
    \sigma[\psi_\tau] \equiv \Delta S[\psi_\tau] +\beta Q[\psi_\tau].
\end{equation}
These definitions are widely used in previous research \cite{horowitz2012quantum,gong2016quantum,hekking2013quantum,liu2016calculating}.
The average values of these quantities agree with the ensemble-level definitions as \begin{math}\langle Q\rangle_s=\langle Q\rangle \end{math} and \begin{math}\langle \Delta S\rangle_s=\langle \Delta S\rangle \end{math}, where \begin{math}\langle \cdot\rangle_s\end{math} denotes ensemble average for the standard trajectories. This agreement guarantees the consistency of the definitions of thermodynamic quantities.

\subsection{Fine unraveling}\label{fine unraveling}
\begin{figure}[tb]
\begin{center}
\includegraphics[width=0.95\textwidth]{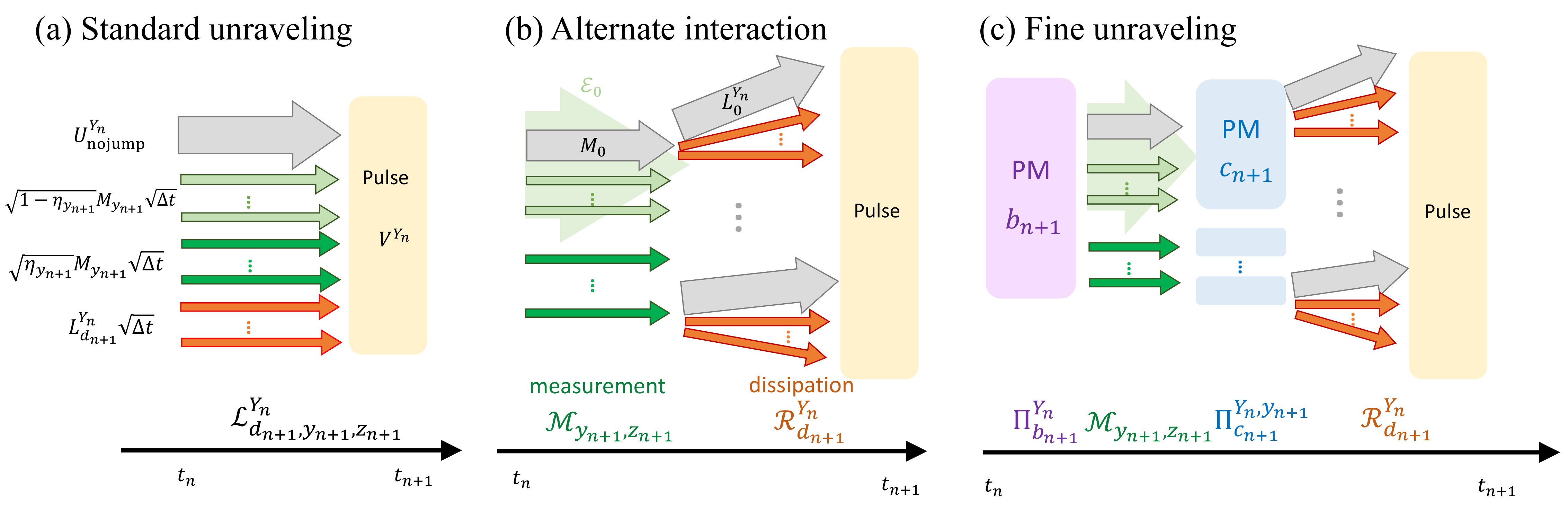}
\caption{Schematics of (a) the standard unraveling, (b) the alternate interaction situation, and (c) the fine unraveling. (a) The set of Kraus operators for the standard unraveling is defined as \eqref{Kraus_Sch_unravel}. (b) The Kraus operators for the alternate interaction situation is the Trotter decomposition of the Kraus operators of the standard unraveling into measurement part and dissipation part. (c) Projective measurements are performed right before and after the separated measurement part. PMs are performed in the diagonal bases of the conditional density operators $\rho_{t_{n}}^{Y_n}$ and $\sigma_{t_{n}}^{Y_{n+1}}$.}
\label{fig:fine_unraveling}
\end{center}
\end{figure}
In this subsection, we first provide the definition of the fine unraveling by introducing its representation by the set of Kraus operators, and ensure that the fine unraveling reproduces the original dynamics of SME \eqref{SME_eta} by taking the ensemble average. We next ensure that the definitions of the stochastic quantities for the fine trajectories (Eqs.~\eqref{letter:QdStraj} and \eqref{stochatic QC transfer entropy} in the main text) are consistent with those at the ensemble level (Eqs.~\eqref{heat_entprd} and \eqref{QC_transfer_entropy} in the main text).

The Kraus operators for the fine unraveling in $[t_n,t_{n+1})$ are defined as 
\begin{equation}
    \label{fine_unraveling_eta}
    \{ \mathcal{R}_{d_{n+1}}^{Y_{n}} \Pi_{c_{n+1}}^{Y_{n},y_{n+1}} \mathcal{M}_{y_{n+1},z_{n+1}}\Pi_{b_{n+1}}^{Y_{n}} \}_{d_{n+1},c_{n+1},y_{n+1},z_{n+1},b_{n+1}},
\end{equation}
where
\begin{equation}
\label{fine_unraveling_eta_detail}
    \begin{split}
    \mathcal{R}_{d_{n+1}}^{Y_n} &\equiv
    \begin{cases}
     V^{Y_n}L_0^{Y_n} &(d_{n+1}=0)\\
     V^{Y_n}L_{d_{n+1}}^{Y_n}\sqrt{\Delta t} &(d_{n+1}\neq0)\\
    \end{cases}\\
    \mathcal{M}_{y_{n+1},z_{n+1}}&\equiv
    \begin{cases}
     M_0 &(y_{n+1} =0,z_{n+1} =0)\\
     \sqrt{\eta_{y_{n+1}}} M_{y_{n+1}}\sqrt{\Delta t} &(y_{n+1} \neq 0,z_{n+1} =y_{n+1})\\
     \sqrt{1-\eta_{z_{n+1}}} M_{z_{n+1}}\sqrt{\Delta t} &(y_{n+1} =0,z_{n+1} \neq 0)\\
    \end{cases}\\
    \rho_{t_n}^{Y_n} &\equiv \sum_{b_{n+1}}p^{Y_n}(b_{n+1}) \Pi_{b_{n+1}}^{Y_n},\\
    \sigma_{t_n}^{Y_{n+1}} &\equiv \sum_{c_{n+1}}p^{Y_{n+1}}(c_{n+1}) \Pi_{c_{n+1}}^{Y_{n+1}},
    \end{split}
\end{equation}
and $\Pi_{b_{n+1}}^{Y_n}$ and $\Pi_{c_{n+1}}^{Y_{n+1}}$ denote the projectors onto the eigenstates of $\rho_{t_n}^{Y_n}$ and $\sigma_{t_n}^{Y_{n+1}}$, respectively.
The conditional density operators right after continuous measurement of $y_{n+1}$ are defined as 
\begin{equation}
    \label{sigmaYt+dt}
    \begin{split}
        \sigma_{t_n}^{Y_n,0} &\equiv \mathcal{E}_0(\rho_{t_n}^{Y_n}) = \mathcal{N}[M_0\rho_{t_n}^{Y_n}M_0^\dag + \sum_{y}(1-\eta_{y})M_{y}\rho_{t_n}^{Y_n}M_{y}^\dag \Delta t],\\
        \sigma_{t_n}^{Y_n,y_{n+1}} &\equiv \mathcal{N}[M_{y_{n+1}}\rho_{t_{n}}^{Y_n}M_{y_{n+1}}^\dag],
    \end{split}
\end{equation}
where $\mathcal{N}[\cdot]$ denotes the normalization of the density operators, and the measurement outcomes of the inserted PMs right before and after the measurement of $y_{n+1}$ are defined as $b_{n+1}$ and $c_{n+1}$.
We further define  $M_0\equiv 1- \frac{1}{2}\{\sum_{y} M_y^\dag M_y\}\Delta t$, and $ L_0^{Y_n}\equiv 1- i(H_{t_n}^{Y_n}+h_{t_n}^{Y_n})\Delta t -\frac{1}{2}\{\sum_d L_d^{Y_n\dag}L_d^{Y_n}\}\Delta t$.
We can see that the alternate interaction situation in Fig.~\ref{fig:unravel} of the main text corresponds to performing the Trotter decomposition of $\mathcal{L}$ to $\mathcal{M}$ and $\mathcal{R}$. The fine unraveling is introduced by further inserting PMs inbetween.
See also Fig.~\ref{fig:fine_unraveling} for the illustration of the fine unraveling.
The fine trajectory until \begin{math}t_n\end{math} is designated by \begin{math}(\psi_{t_n},\pi_{t_n}) = (a_0,D_n,Z_n,Y_n,\pi_{t_n})\end{math}, where we introduce $\pi_{t_n} \equiv\{ (b_1,b_2,\dots,b_n),(c_1,c_2,\dots,c_n)\}$ to denote all the results of the inserted PMs.

We can confirm that the set of Kraus operators \eqref{fine_unraveling_eta} reproduces the SME~\eqref{SME_eta} by taking the ensemble average of \begin{math}(a_0,D_n,Z_n,\pi_{t_n})\end{math} in the limit of \begin{math}\Delta t \to 0\end{math}, because
\begin{equation}
\label{fine_unraveling_condition}
    \mathbb{E}_{Y_n}[\ket{\psi_{t_n},\pi_{t_n}}\bra{\psi_{t_n},\pi_{t_n}}] \equiv \sum_{(\psi_{t_n},\pi_{t_n})|Y_n} P_f[\psi_{t_n},\pi_{t_n}] \ket{\psi_{t_n},\pi_{t_n}}\bra{\psi_{t_n},\pi_{t_n}}= P[Y_n]\rho_{t_n}^{Y_n}(1+\mathcal{O}(\Delta t))
\end{equation}
holds. Here, \begin{math}\sum_{(\psi_{t_n},\pi_{t_n})|Y_n}\end{math} is defined as the summation over all the fine trajectories under fixed \begin{math}Y_n\end{math}. \begin{math}P_f[\psi_{t_n},\pi_{t_n}]\end{math} is the probability that takes the fine trajectory \begin{math}(\psi_{t_n},\pi_{t_n})\end{math} until time \begin{math}t_n\end{math} and \begin{math}\ket{\psi_{t_n},\pi_{t_n}}\end{math} is the normalized state:
\begin{equation}
    \label{fine traj prob def}
    P_f[\psi_{t_n},\pi_{t_n}] \equiv \|\mathcal{R}^{Y_{n-1}}_{d_n}\Pi_{c_n}^{Y_{n-1},y_n}\mathcal{M}_{y_n,z_n}\Pi_{b_n}^{Y_{n-1}} \dots\Pi_{c_1}^{Y_1}\mathcal{M}_{y_1,z_1}\Pi_{b_1}\ket{a_0}\|^2,
\end{equation}
\begin{equation}
    \label{fine traj ket def}
    \ket{\psi_{t_n},\pi_{t_n}} \equiv \mathcal{N}[\mathcal{R}^{Y_{n-1}}_{d_n,z_n}\Pi_{c_n}^{Y_{n-1},y_n}\mathcal{M}_{y_n,z_n}\Pi_{b_n}^{Y_{n-1}} \dots\Pi_{c_1}^{Y_1}\mathcal{M}_{y_1,z_1}\Pi_{b_1}\ket{a_0}].
\end{equation}
We abbreviate the subscript $f$ of \begin{math}P_f\end{math} in obvious cases.
Equation~\eqref{fine_unraveling_condition} can be inductively derived by using Eqs.~\eqref{fine_un_calc} and \eqref{fine_un_calc2} below, where we calculate the ensemble average over \begin{math}c_{n+1},b_{n+1},z_{n+1},d_{n+1}\end{math} with fixed $y_{n+1}$ (we suppose $y_{n+1} \neq0$ in Eq.~\eqref{fine_un_calc} and $y_{n+1} = 0$ in Eq.~\eqref{fine_un_calc2}.):
\begin{equation}
\label{fine_un_calc}
\begin{split}
    &\sum_{c,b,z=y_{n+1},d} \mathcal{R}_{d}^{Y_{n}} \Pi_{c}^{Y_{n},y_{n+1}} \mathcal{M}_{y_{n+1},z}\Pi_{b}^{Y_{n}} \rho_{t_n}^{Y_n}\Pi_{b}^{Y_{n}}\mathcal{M}_{y_{n+1},z}^{\dag}\Pi_{c}^{Y_{n},y_{n+1}}\mathcal{R}_{d}^{Y_n \dag} \\
    =&\sum_{c,b,d}\eta_{y_{n+1}} \mathcal{R}_{d}^{Y_{n}}\Pi_{c}^{Y_{n},y_{n+1}}M_{y_{n+1}}\Pi_{b}^{Y_n}\rho_{t_n}^{Y_n}\Pi_{b}^{Y_n}M_{y_{n+1}}^\dag\Pi_{c}^{Y_{n},y_{n+1}}\mathcal{R}_{d}^{Y_{n}\dag} \Delta t \\
    =&P[y_{n+1}|Y_n] \{\sum_{d}\mathcal{R}_{d}^{Y_{n}}\sigma_{t_{n}}^{Y_n,y_{n+1}}\mathcal{R}_{d}^{Y_{n}\dag}\}=  P[y_{n+1}|Y_n]\rho_{t_{n+1}}^{Y_n,y_{n+1}} (1+\mathcal{O}(\Delta t)),\\
\end{split}
\end{equation}
\begin{equation}
    \label{fine_un_calc2}
    \begin{split}
        \sum_{c,b,z,d}& \mathcal{R}_{d}^{Y_{n}} \Pi_{c}^{Y_{n},0} \mathcal{M}_{0,z}\Pi_{b}^{Y_{n}} \rho_{t_n}^{Y_n}\Pi_{b}^{Y_{n}}\mathcal{M}_{0,z}^{\dag}\Pi_{c}^{Y_{n},0}\mathcal{R}_{d}^{Y_n \dag} \\
        =\sum_{c,b}& V^{Y_n}L_0^{Y_n}\Pi_{c}^{Y_{n},0}M_0\Pi_{b}^{Y_n}\rho_{t_n}^{Y_n}\Pi_{b}^{Y_n}M_0^\dag\Pi_{c}^{Y_{n},0}L_0^{Y_n\dag} V^{Y_n\dag}
        + \sum_{d\neq0,c,b} V^{Y_n}L_{d}^{Y_n}\Pi_{c}^{Y_{n},0}M_0\Pi_{b}^{Y_n}\rho_{t_n}^{Y_n}\Pi_{b}^{Y_n}M_0^\dag\Pi_{c}^{Y_{n},0}L_{d}^{Y_n\dag}V^{Y_n\dag} \Delta t \\
        &+\sum_{z\neq0,c,b,d} (1-\eta_{z})\mathcal{R}_{d}^{Y_{n}}\Pi_{c}^{Y_{n},0}M_{z}\Pi_{b}^{Y_n}\rho_{t_n}^{Y_n}\Pi_{b}^{Y_n}M_{z}^\dag\Pi_{c}^{Y_{n},0}\mathcal{R}_{d}^{Y_{n}\dag} \Delta t\\
        =\sum_{c}& V^{Y_n}L_0^{Y_n}\Pi_{c}^{Y_{n},0}M_0\rho_{t_n}^{Y_n}M_0^\dag\Pi_{c}^{Y_{n},0}L_0^{Y_n\dag}V^{Y_n\dag}
        + \sum_{d\neq0,c} V^{Y_n}L_{d}^{Y_n}\Pi_{c}^{Y_{n},0}M_0\rho_{t_n}^{Y_n}M_0^\dag\Pi_{c}^{Y_{n},0}L_{d}^{Y_n\dag}V^{Y_n\dag} \Delta t \\
        &+\sum_{z\neq0,c,d} (1-\eta_{z})\mathcal{R}_{d}^{Y_{n}}\Pi_{c}^{Y_{n},0}M_{z}\rho_{t_n}^{Y_n}M_{z}^\dag\Pi_{c}^{Y_{n},0}\mathcal{R}_{d}^{Y_n \dag} \Delta t\\
        =\sum_{c}& V^{Y_n}L_0^{Y_n}\Pi_{c}^{Y_{n},0}\bigl(M_0\rho_{t_n}^{Y_n}M_0^\dag + \sum_{z\neq0}(1-\eta_{z})M_{z}\rho_{t_n}^{Y_n}M_{z}^\dag \Delta t\bigr)\Pi_{c}^{Y_{n},0}L_0^{Y_n\dag} V^{Y_n\dag}\\
        &+ \sum_{d\neq0,c} V^{Y_n}L_{d}^{Y_n}\Pi_{c}^{Y_{n},0}\bigl(M_0\rho_{t_n}^{Y_n}M_0^\dag + \sum_{z\neq0}(1-\eta_{z})M_{z}\rho_{t_n}^{Y_n}M_{z}^\dag \Delta t\bigr)\Pi_{c}^{Y_{n},0}L_{d}^{Y_n\dag} V^{Y_n\dag}\Delta t \\
        =\sum_{c}& V^{Y_n}L_0^{Y_n}\Pi_{c}^{Y_{n},0}P[0|Y_n]\sigma_{t_n}^{Y_n,0}\Pi_{c}^{Y_{n},0}L_0^{Y_n\dag}V^{Y_n\dag} + \sum_{d\neq0,c} V^{Y_n}L_{d}^{Y_n}\Pi_{c}^{Y_{n},0}P[0|Y_n]\sigma_{t_n}^{Y_n,0}\Pi_{c}^{Y_{n},0}L_{d}^{Y_n\dag}V^{Y_n\dag} \Delta t \\
        = P[0&|Y_n] V^{Y_n}\Bigl\{L_0^{Y_n}\sigma_{t_n}^{Y_n,0}L_0^{Y_n\dag}+ \sum_{d\neq0} L_{d}^{Y_n}\sigma_{t_n}^{Y_n,0}L_{d}^{Y_n\dag} \Delta t\Bigr\}V^{Y_n\dag}  =P[0|Y_n]\rho_{t_{n+1}}^{Y_{n},0}(1+\mathcal{O}(\Delta t^2)).\\
    \end{split}
\end{equation}
We note that the inserted PMs must be performed in the diagonal bases of the conditional density operators \begin{math}\rho_{t_n}^{Y_n},\sigma_{t_n}^{Y_{n+1}}\end{math}. Otherwise, the set of Kraus operators is not an unraveling of SME~\eqref{SME_eta}, because we cannot reproduce the dynamics of the SME \eqref{SME_eta} by taking the ensemble average over trajectories.

The thermodynamic quantities can be defined to each fine trajectory in the same way as is the case for the standard trajectory (Eq.~\eqref{Q_Sch}, \eqref{dS_Sch}, and \eqref{sig_Sch}).
We can make sure that the definition of dissipated heat \begin{math}Q[\psi_\tau,\pi_\tau]\end{math} to each fine trajectory (Eq.~\eqref{Q_Sch}) is consistent with the ensemble-level definition (Eq.\eqref{Q_ens}) in the continuous time limit $\Delta t\to 0$ as
\begin{equation}
    \label{Q_fine_verify}
    \begin{split}
        \langle Q\rangle_f=&\sum_{\psi_\tau,\pi_\tau}P_f[\psi_\tau,\pi_\tau]Q[\psi_\tau,\pi_\tau] \\
        =& \sum_{\psi_\tau,\pi_\tau}P_f[\psi_\tau,\pi_\tau]\sum_{n=0}^{N-1}\sum_{d_{n+1}} \Delta N_{d_{n+1}}^{Y_n} \Delta_{d_{n+1}}^{Y_n}\\
        =&\sum_{n=0}^{N-1}\sum_{\psi_\tau,\pi_\tau} \sum_{d_{n+1}} P_f[\psi_\tau,\pi_\tau] \Delta N_{d_{n+1}}^{Y_n} \Delta_{d_{n+1}}^{Y_n} \\
        =&\sum_{n=0}^{N-1}\sum_{\psi_{t_n},\pi_{t_n}} \sum_{b_{n+1},y_{n+1},z_{n+1},c_{n+1},d_{n+1}} P_f[\psi_{t_n},\pi_{t_n}] \|L_{d_{n+1}}^{Y_n}\Pi_{c_{n+1}}^{Y_{n},y_{n+1}} \mathcal{M}_{y_{n+1},z_{n+1}}\Pi_{b_{n+1}}^{Y_{n}}\ket{\psi_{t_n},\pi_{t_n}}\|^{2}\Delta t \Delta_{d_{n+1}}^{Y_n}  \\
        =&\sum_{n=0}^{N-1} \sum_{Y_{n}}\sum_{d_{n+1}} \tr[L_{d_{n+1}}^{Y_n}\bigl\{P[Y_n,y_{n+1}=0]\sigma_{t_n}^{Y_n,0}+\sum_{y_{n+1}\neq0} P[Y_n,y_{n+1}]\sigma_{t_n}^{Y_n,y_{n+1}}\bigr\}L_{d_{n+1}}^{Y_n\dag}] (1+\mathcal{O}(\Delta t))\Delta t  \Delta_{d_{n+1}}^{Y_n}  \\
        =&\sum_{n=0}^{N-1} \sum_{Y_n}\sum_{d_{n+1}} P[Y_n]\tr[L_{d_{n+1}}^{Y_n}\rho_{t_n}^{Y_n}L_{d_{n+1}}^{Y_n\dag}](1+\mathcal{O}(\Delta t))\Delta t \Delta_{d_{n+1}}^{Y_n}   = \langle Q\rangle (1+\mathcal{O}(\Delta t)),\\
    \end{split}
\end{equation}
where \begin{math}\langle \cdot\rangle_f\end{math} denotes ensemble average for the fine trajectories. 
We used Eq.~\eqref{fine_unraveling_condition} to obtain the fifth line.
We can confirm that the other thermodynamic quantities also fulfill such consistency. 

We can also show that the ensemble average of the stochastic QC-transfer entropy \begin{math}i_{\mathrm{QC}}[\psi_\tau,\pi_\tau]\end{math} (Eq.~\eqref{stochatic QC transfer entropy}) coincides with $\langle i_{\mathrm{QC}} \rangle$ (Eq.~\eqref{QC_transfer_entropy}) in the limit of $\Delta t\to 0$ as
\begin{equation}
    \label{QC_verify}
    \begin{split}
        \langle i_{\mathrm{QC}} \rangle_f &= \sum_{\psi_\tau,\pi_\tau}P_f[\psi_\tau,\pi_\tau]i_{\mathrm{QC}}[\psi_\tau,\pi_\tau] \\
        &= \sum_{\psi_\tau,\pi_\tau}P_f[\psi_\tau,\pi_\tau]\sum_{n=0}^{N-1} -\ln p^{Y_n}(b_{n+1}) +\ln p^{Y_{n+1}}(c_{n+1}) \\
        &= \sum_{n=0}^{N-1}\Big\{\sum_{Y_n,b_{n+1}}\sum_{(\psi_{t_n},\pi_{t_n})|Y_n}-P_f[\psi_{t_n},\pi_{t_n},b_{n+1}] \ln p^{Y_n}(b_{n+1})\\ 
        &\ \ \ \ \ \ \ +\sum_{Y_{n},y_{n+1},z_{n+1},c_{n+1},b_{n+1}}\sum_{(\psi_{t_n},\pi_{t_n})|Y_n} P_f[\psi_{t_n},\pi_{t_{n+1}},y_{n+1},z_{n+1}] \ln p^{Y_{n},y_{n+1}}(c_{n+1}) \Big\} \\
        &= \sum_{n=0}^{N-1}\Big\{\sum_{Y_n,b_{n+1}}-P[Y_n] \tr[\Pi_{b_{n+1}}^{Y_n}\rho_{t_n}^{Y_n}\Pi_{b_{n+1}}^{Y_n}] \ln p^{Y_n}(b_{n+1})\\
        &\ \ \ \ \ \ \  +\sum_{Y_{n},y_{n+1},c_{n+1}}P[Y_n] \tr[\Pi_{c_{n+1}}^{Y_n,y_{n+1}}P[y_{n+1}|Y_n] \sigma_{t_n}^{Y_n,y_{n+1}}\Pi_{c_{n+1}}^{Y_n,y_{n+1}}] \ln p^{Y_n,y_{n+1}}(c_{n+1}) \Big\} (1+\mathcal{O}(\Delta t))\\
        &= \sum_{n=0}^{N-1}\sum_{Y_n}P[Y_n] \Big\{\sum_{b_{n+1}}- p^{Y_n}(b_{n+1})\ln p^{Y_n}(b_{n+1}) \\
        &\ \ \ \ \ \ \ + \sum_{c_{n+1},y_{n+1}}P[y_{n+1}|Y_n]p^{Y_n,y_{n+1}}(c_{n+1}) \ln p^{Y_n,y_{n+1}}(c_{n+1}) \Big\}(1+\mathcal{O}(\Delta t)) \\
        &= \Bigg[\sum_{n=0}^{N-1}\sum_{Y_n}P[Y_n] \Big\{S(\rho_{t_n}^{Y_n}) -\sum_{y_{n+1}}P[y_{n+1}|Y_n] S(\sigma_{t_{n}}^{Y_n,y_{n+1}}) \Big\}\Bigg](1+\mathcal{O}(\Delta t))\\
        &= \langle i_{\mathrm{QC}} \rangle (1+\mathcal{O}(\Delta t)). \\
    \end{split}
\end{equation}
We again used Eq.~\eqref{fine_unraveling_condition} to obtain the fourth line.

As can be seen from Eq.~\eqref{QC_verify}, the QC-transfer entropy can be interpreted as the accumulation of conditional QC-mutual information \cite{ozawa1986information,groenewold1971problem} if continuous measurement is perfect (i.e., $\eta_y=1$). If continuous measurement is imperfect (i.e., $\eta_y<1$), the increment of QC-transfer entropy is replaced by the ``information gain", introduced in Ref.~\cite{funo2013integral}.

\section{derivation of the main results}\label{derivation of main results}
In this section, we prove the main result of this work: the FT \eqref{generalized_fluctuation_theorem} in the main text. In Section \ref{Overview of the derivation}, we show the overview of the derivation. In Section \ref{Derivation of the GDFT under fine unraveling} and \ref{Absolute irreversibility of generalized FT in fine unraveling}, the details of the proof are provided.
\subsection{Overview of the derivation}\label{Overview of the derivation}
\begin{figure}[tb]
\begin{center}
\includegraphics[width=0.8\textwidth]{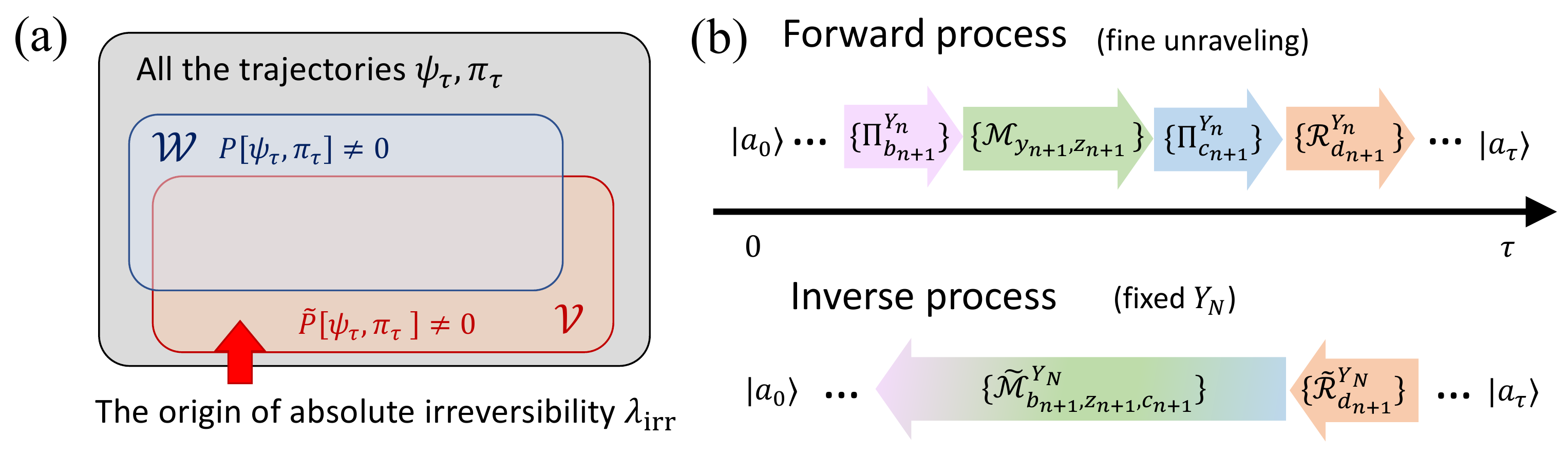}
\caption{(a) Venn diagram of the fine trajectory. The fine trajectories that satisfy \begin{math}P[\psi_\tau,\pi_\tau] \neq 0\end{math} are in \begin{math}\mathcal{W}\end{math}. The GDFT holds only for the trajectories in this set. The trajectories satisfying \begin{math}\Tilde{P}[\psi_\tau,\pi_\tau] \neq 0\end{math} are in \begin{math}\mathcal{V}\end{math}. Absolute irreversibility term in the generalized FT \begin{math}\lambda_{\mathrm{irr}}\end{math} is the probability sum of all the trajectories in \begin{math}\mathcal{V}\smallsetminus\mathcal{W}\end{math}. (b) Forward and inverse processes for the fine unraveling. The forward process is defined as Fig.~\ref{fig:fine_unraveling}. The corresponding inverse process is defined as the consecutive composition of CPTP maps (defined as \eqref{Kraus_inverse_fine_detail_M} and \eqref{Kraus_inverse_fine_detail_R}) in the inverse order. The trace-preserving nature of the inverse process guarantees the probability normalization condition Eq.~\eqref{normalization condition}.}
\label{fig:deriv_generalized FT}
\end{center}
\end{figure}
The essence of the derivation of the generalized FT is to introduce the inverse process. We define inverse trajectories so that they have natural correspondence to forward trajectories. Here, forward process and trajectory mean the dynamics that follow SME~\eqref{SME_eta} and its unraveled trajectory, respectively.
We impose two requirements upon the inverse process. Namely, the inverse process must satisfy the generalized detailed fluctuation theorem (GDFT) Eq.~\eqref{Dgeneralized FT} and probability normalization condition Eq.~\eqref{normalization condition}:
\begin{equation}
    \label{Dgeneralized FT}
    \frac{\Tilde{P}_f[\psi_\tau,\pi_\tau]}{P_f[\psi_\tau,\pi_\tau]} = e^{-\sigma[\psi_\tau,\pi_\tau]-i_{\mathrm{QC}}[\psi_\tau,\pi_\tau]},
\end{equation}
\begin{equation}
    \label{normalization condition}
\sum_{\psi_\tau,\pi_\tau} \Tilde{P}_f[\psi_\tau,\pi_\tau]= 1.
\end{equation}
Here, $\Tilde{P}_f$ denotes the probability of the inverse trajectory.
The GDFT (Eq.~\eqref{Dgeneralized FT}) is the equation about the probability ratio of the corresponding forward and inverse trajectories. The GDFT only holds in the realizable forward trajectories (\begin{math}P_f[\psi_\tau,\pi_\tau] \neq0\end{math}) since denominator must be nonzero.

By utilizing this inverse process, we can derive the (integral) generalized FT as 
\begin{equation}
\label{generalized FTderv_overview}
\begin{split}
\langle e^{-\sigma-i_{\mathrm{QC}}} \rangle &=\sum_{(\psi_\tau,\pi_\tau)\in \mathcal{W}} P_f[\psi_\tau,\pi_\tau] \frac{\Tilde{P}_f[\psi_\tau,\pi_\tau]}{P_f[\psi_\tau,\pi_\tau]} \\
&= \sum_{(\psi_\tau,\pi_\tau)\in \mathcal{W}}\Tilde{P}_f[\psi_\tau,\pi_\tau]=1-\lambda_{\mathrm{irr}},
\end{split}
\end{equation}
where
\begin{equation}
\label{absirr_ov}
    \lambda_{\mathrm{irr}} \equiv \sum_{(\psi_\tau,\pi_\tau)\in (\mathcal{V}\smallsetminus \mathcal{W})}\Tilde{P}_f[\psi_\tau,\pi_\tau].
\end{equation}
Here, \begin{math}\mathcal{W}\end{math} denotes the set of forward trajectories whose realization probabilities are nonzero (i.e., $P_f[\psi_\tau,\pi_\tau]\neq0$), and \begin{math}\mathcal{V}\end{math} denotes the set of forward trajectories which satisfy $\Tilde{P}_f[\psi_\tau,\pi_\tau]\neq0$ (see Fig.~\ref{fig:deriv_generalized FT}(a)). Absolute irreversibility term (Eq.~\eqref{absirr_ov}) arises from the gap between realizable forward trajectory and inverse trajectory. This gap is nonzero when the measurement $\{M_{y_n}\}$ works as the projection onto the Hilbert subspace. Such absolute irreversibility term is also present in single quantum measurement case \cite{funo2015quantum,murashita2017fluctuation}. We derive a sufficient condition for \begin{math}\lambda_{\mathrm{irr}} =0\end{math} in Section \ref{Absolute irreversibility of generalized FT in fine unraveling}.

The inverse process introduced here cannot be interpreted as the ordinary time-reversal of the forward process as in the conventional detailed FT \cite{funo2018quantum}. This is because it is difficult to devise the time-reversal of quantum measurement. Therefore, we have to devise a kind of artificial inverse process for the GDFT, which can be regarded as a useful tool to derive the (integral) generalized FT.

\subsection{Derivation of the GDFT under the fine unraveling}\label{Derivation of the GDFT under fine unraveling}
In this subsection, we define the inverse process for the fine unraveling, and derive the GDFT. Because of the presence of feedback, we define the inverse process depending on the measurement outcomes \begin{math}Y_N\end{math}. The realization probability of each inverse trajectory is defined as
\begin{equation}
    \label{inverse_traj_prob_Y}
    \Tilde{P}_f[\psi_\tau,\pi_\tau] \equiv P[Y_N]\ \Tilde{p}_f^{Y_N}[\psi_\tau,\pi_\tau],
\end{equation} 
where \begin{math}\Tilde{p}_f^{Y_N}\end{math} is defined as
\begin{equation}
    \label{inverse_prob_fine}
    \Tilde{p}_f^{Y_N}[\psi_\tau,\pi_\tau] \equiv \|\bra{a_0}\Tilde{\mathcal{M}}_{b_{1},z_{1},c_{1}}^{Y_N}\Tilde{\mathcal{R}}_{d_{1}}^{Y_N}\Tilde{\mathcal{M}}_{b_{2},z_{2},c_{2}}^{Y_N}\Tilde{\mathcal{R}}_{d_{2}}^{Y_N} \dots \Tilde{\mathcal{M}}_{b_{N},z_{N},c_{N}}^{Y_N}\Tilde{\mathcal{R}}_{d_{N}}^{Y_N}\ket{a_\tau}\|^2p_\tau (a_\tau).
\end{equation}
Here, \begin{math}\Tilde{\mathcal{M}}_{b_{n},z_{n},c_{n}}^{Y_N}\end{math} represents the inverse process of the measurement and PMs \begin{math}\{\Pi_{c_{n}}^{Y_{n-1},y_{n}} \mathcal{M}_{y_{n},z_{n}}\Pi_{b_{n}}^{Y_{n-1}}\}\end{math}, and \begin{math}\Tilde{\mathcal{R}}_{d_{n}}^{Y_N}\end{math} denotes the inverse of the heat-bath dissipation \begin{math}\{\mathcal{R}_{d_{n}}^{Y_{n-1}}\}\end{math}. They are defined as
\begin{equation}
\label{Kraus_inverse_fine_detail_M}
\Tilde{\mathcal{M}}_{b_{n},z_{n},c_{n}}^{Y_N}\equiv
\begin{cases}
      \ket{b_{n}}^{Y_{n-1}}{}^{Y_{n-1},y_{n}}\bra{c_{n}}\sqrt{\frac{\|{}^{Y_{n-1},y_{n}}\!\bra{c_{n}}\mathcal{M}_{y_{n},z_{n}}\ket{b_{n}}^{Y_{n-1}}\|^2 p^{Y_{n-1}}(b_{n})}{p^{Y_{n-1},y_{n}}(c_{n})P[y_{n}|Y_{n-1}]}}
      &(P^{Y_{n-1},y_{n}}(c_{n})P[y_{n}|Y_{n-1}] \neq 0)\\
      \ket{b_{n}}^{Y_{n-1}}{}^{Y_{n-1},y_{n}}\bra{c_{n}} \frac{1}{\sqrt{\alpha}}&(P^{Y_{n-1},y_{n}}(c_{n})P[y_{n}|Y_{n-1}]  = 0),\\
\end{cases}
\end{equation}
\begin{equation}
\label{Kraus_inverse_fine_detail_R}
    \Tilde{\mathcal{R}}_{d_{n}}^{Y_N} \equiv
    \begin{cases}
     L_0^{Y_{n-1}\dag}V^{Y_{n-1}\dag} &(d_{n} =0)\\
     L_{d^{\prime}}^{Y_{n-1}}\sqrt{\Delta t}V^{Y_{n-1}\dag} &(d_{n}\neq0),\\
    \end{cases}
\end{equation}
where \begin{math}\alpha\end{math} is the Hilbert space dimension of the system and $L_{d^{\prime}}^{Y_{n-1}}$ is the heat-bath dissipation operator that satisfies the detailed balance condition $L_{d^{\prime}}^{Y_{n-1}\dag}=L_{d_{n}}^{Y_{n-1}}e^{-\beta\frac{\Delta_{d_{n}}}{2}}$.
Here, the diagonal bases $\{\ket{b_{n}}^{Y_{n-1}}\}, \{\ket{c_{n}}^{Y_{n}}\}$ of the conditional density operators $\rho_{t_n-1}^{Y_{n-1}}$ and $\sigma_{t_{n-1}}^{Y_{n}}$ are defined as $\rho_{t_{n-1}}^{Y_{n-1}}\equiv \sum_{b_{n}}p^{Y_{n-1}}(b_{n})\ket{b_{n}}^{Y_{n-1}} {}^{Y_{n-1}}\!\bra{b_{n}}$ and $\sigma_{t_n-1}^{Y_{n}}\equiv \sum_{c_{n}}p^{Y_{n}}(c_{n})\ket{c_{n}}^{Y_{n}}  {}^{Y_{n}}\!\bra{c_{n}}$, respectively. We abbreviate the superscript $Y_n$ of the diagonalized bases in obvious cases.
The operators \eqref{Kraus_inverse_fine_detail_M}, \eqref{Kraus_inverse_fine_detail_R} fulfill the completeness condition \begin{math}\sum_{b_{n},z_{n},c_{n},d_{n}} \Tilde{\mathcal{R}}_{d_{n}}^{Y_N\dag}\Tilde{\mathcal{M}}_{b_{n},z_{n},c_{n}}^{Y_N\dag}\Tilde{\mathcal{M}}_{b_{n},z_{n},c_{n}}^{Y_N}\Tilde{\mathcal{R}}_{d_{n}}^{Y_N} =1\end{math}. Therefore, the inverse process is defined as the consecutive composition of CPTP maps in the inverse order (see Fig.~\ref{fig:deriv_generalized FT}(b)).
Because of this completeness condition, the probability normalization condition holds, i.e., \begin{math}\sum_{\psi_\tau,\pi_\tau}\Tilde{P}_f[\psi_\tau,\pi_\tau]= \sum_{Y_N}P[Y_N]\sum_{(\psi_\tau,\pi_\tau)|Y_N} \Tilde{p}_f^{Y_N}[\psi_\tau,\pi_\tau] =1\end{math}.

We can verify the GDFT (Eq.\eqref{Dgeneralized FT}) for \begin{math}(\psi_\tau,\pi_\tau) \in \mathcal{W}\end{math} as
\begin{equation}
    \label{Dgeneralized FT_derivation}
    \begin{split}
        \frac{\Tilde{P}_f[\psi_\tau,\pi_\tau]}{P_f[\psi_\tau,\pi_\tau]} &=\frac{P[Y_N]\ \Tilde{p}_f^{Y_N}[\psi_\tau,\pi_\tau]}{P_f[\psi_\tau,\pi_\tau]} \\
        &=P[Y_N]e^{-\Delta S}\frac{\| \bra{a_\tau}\Tilde{\mathcal{R}}_{d_{N}}^{Y_N\dag} \ket{c_{N}}\bra{c_{N}}\Tilde{\mathcal{M}}_{b_{N},z_N,c_{N}}^{Y_N\dag}\ket{b_{N}}\bra{b_{N}}\dots \bra{c_{1}} \Tilde{\mathcal{M}}_{b_{1},z_{1},c_{1}}^{Y_N\dag}\ket{b_1} \|^2}
        {\|\bra{a_\tau}\mathcal{R}_{d_{N}}^{Y_{N-1}} \ket{c_{N}}\bra{c_{N}} \mathcal{M}_{y_{N},z_{N}}\ket{b_{N}}\bra{b_{N}} \dots \bra{c_{1}} \mathcal{M}_{y_{1},z_{1}}\ket{b_1}\|^2}\\
        &=P[Y_N]e^{-\Delta S}
        \frac{\|\bra{a_\tau}\Tilde{\mathcal{R}}_{d_{N}}^{Y_N\dag} \ket{c_{N}}\|^2}{\|\bra{a_\tau}\mathcal{R}_{d_{N}}^{Y_{N-1}} \ket{c_{N}}\|^2}\frac{\|\bra{c_{N}}\Tilde{\mathcal{M}}_{b_{N},z_N,c_{N}}^{Y_N\dag}\ket{b_{N}}\|^2}{\|\bra{c_{N}} \mathcal{M}_{y_{N},z_{N}}\ket{b_{N}}\|^2}  \dots
        \frac{\|\bra{c_{1}} \Tilde{\mathcal{M}}_{b_1,z_1,c_1}^{Y_N\dag}\ket{b_1}\|^2}{\|\bra{c_{1}} \mathcal{M}_{y_{1},z_{1}}\ket{b_1}\|^2}\\
        &=P[Y_N]e^{-\Delta S}
        \exp{\Bigl(-\beta\sum_{d_N}\Delta_{d_{N}}^{Y_{N-1}}\Delta N_{d_{N}}^{Y_{N-1}}\Bigr)} \frac{p^{Y_{N-1}}(b_{N})}{p^{Y_N}(c_{N})P[y_{N}|Y_{N-1}]}\dots \frac{p^{Y_0}(b_1)}{p^{Y_{1}}(c_1)P[y_1]}\\
        &=P[Y_N]  e^{-\Delta S}\exp{\Bigl(-\beta\sum_{n=0}^{N-1}\sum_{d_{n+1}} \Delta N_{d_{n+1}}^{Y_{n}}\Delta_{d_{n+1}}^{Y_{n}}\Bigr)}\cdot\Biggl[\prod_{n=0}^{N-1}\frac{p^{Y_n}(b_{n+1})}{p^{Y_{n+1}}(c_{n+1})P[y_{n+1}|Y_n]}\Biggr]\\
        &=e^{-(\Delta S+\beta Q)} \Biggl[\prod_{n=0}^{N-1}\frac{p^{Y_n}(b_{n+1})}{p^{Y_{n+1}}(c_{n+1})}\Biggr] =e^{-\sigma[\psi_\tau,\pi_\tau]-i_{\mathrm{QC}}[\psi_\tau,\pi_\tau]}.\\
    \end{split}
\end{equation}
We note that \begin{math}\Tilde{\mathcal{M}}_{b_{n+1},z_{n+1},c_{n+1}}^{Y_N}\end{math} is the first line of \eqref{Kraus_inverse_fine_detail_M} for all \begin{math}n\end{math} since \begin{math}p^{Y_n,y_{n+1}}(c_{n+1})P[y_{n+1}|Y_n]\neq 0\end{math} always holds for trajectories in \begin{math}\mathcal{W}\end{math}.
We used \begin{math}\Tilde{\mathcal{M}}_{b_{n+1},z_{n+1},c_{n+1}}^{Y_N}\Tilde{\mathcal{R}}_{d_{n+1}}^{Y_N} =\Pi_{c_{n+1}}^{Y_{n+1}} \Tilde{\mathcal{M}}_{b_{n+1},z_{n+1},c_{n+1}}^{Y_N}\Pi_{b_{n+1}}^{Y_n}\Tilde{\mathcal{R}}_{d_{n+1}}^{Y_N}\end{math} to obtain the second line. From the third to fourth line, we used the detailed balance condition. 

\subsection{Absolute irreversibility term in the generalized FT under the fine unraveling} \label{Absolute irreversibility of generalized FT in fine unraveling}
In this subsection, we show that the \emph{full rank condition} is a sufficient condition for \begin{math}\lambda_{\mathrm{irr}}=0\end{math}, where the condition means that
\begin{equation}
    \label{fullrank_cond}
    \det[\rho_0 ] \neq 0 \  \mathrm{and}\  \det[M_{y_n}] \neq 0\ \mathrm{for\ all\ }y_n.
\end{equation}
We prove the sufficiency by focusing on trajectories in \begin{math}\mathcal{V}\smallsetminus \mathcal{W}\end{math}, and show the contraposition (i.e., \begin{math}\lambda_{\mathrm{irr}} \neq 0\end{math} implies that \begin{math}\det[\rho_0]=0\end{math} or \begin{math}\det[M_{y_n}]=0\end{math} for some $y_n$).

First of all, we need the following three formulas for the proof.
\begin{equation}
\label{prop_absirr}
\begin{split}
    \|\bra{b_{n+1}}\mathcal{R}_{d_{n}}^{Y_{n-1}} \ket{c_{n}}\|^2=0 &\iff \|\bra{b_{n+1}}\Tilde{\mathcal{R}}_{d_{n}}^{Y_N\dag}\ket{c_n}\|^2=0;\\
    \mathrm{if}\ p^{Y_{n-1},y_{n}}(c_{n})P[y_{n}|Y_{n-1}] \neq 0,\   \|\bra{c_{n}} \mathcal{M}_{y_{n},z_{n}}\ket{b_{n}}\|^2 &= 0 \ \mathrm{implies}\  \|\bra{c_{n}} \Tilde{\mathcal{M}}_{b_{n},z_{n},c_{n}}^{Y_N\dag}\ket{b_{n}} \|^2=0;\\
    \ p_0(a_0) = 0 \ \mathrm{for\ some\ }a_0&\iff \det[\rho_0]=0.\\
\end{split}
\end{equation}
We can show the first line of \eqref{prop_absirr} using the detailed balance condition imposed for heat-bath dissipation, and the second line by the definition of \begin{math}\Tilde{\mathcal{M}}_{b_{n},z_{n},c_{n}}^{Y_N\dag}\end{math} \eqref{Kraus_inverse_fine_detail_M}. The third line follows from the diagonalization of the initial density operator \begin{math}\rho_0 \equiv \sum_{a_0} p_0(a_0)\ket{a_0}\bra{a_0}\end{math}.

The realization probabilities of forward trajectory and the corresponding inverse trajectory are given as
\begin{equation}
    \label{for_inv_prob}
    \begin{split}
    P_f[\psi_\tau,\pi_\tau]&=\|\bra{a_\tau}\mathcal{R}_{d_{N}}^{Y_{N}} \ket{c_N}\bra{c_N} \mathcal{M}_{y_{N},z_{N}}\ket{b_{N}} \dots \bra{c_{1}} \mathcal{M}_{y_{1},z_{1}}\ket{b_1}\langle b_1|a_0\rangle\|^2\cdot p_0(a_0), \\
    \Tilde{P}_f[\psi_\tau,\pi_\tau] &= \| \bra{a_\tau}\Tilde{\mathcal{R}}_{d_{N}}^{Y_N\dag} \ket{c_N}\bra{c_N}\Tilde{\mathcal{M}}_{b_{N},z_{N},c_N}^{Y_N\dag}\ket{b_{N}}\dots \bra{c_{1}} \Tilde{\mathcal{M}}_{b_1,z_1,c_1}^{Y_N\dag}\ket{b_1} \langle b_1|a_0\rangle \|^2 P[Y_N]p_\tau(a_\tau).
   \end{split}
\end{equation}
As can be seen from~\eqref{prop_absirr} and Eq.~\eqref{for_inv_prob}, absolute irreversibility can only arise from the singularity of the initial state, \begin{math}\det[\rho_0] =0\end{math}, or irreversibility of the measurement, \begin{math}p^{Y_n,y_{n+1}}(c_{n+1})P[y_{n+1}|Y_n] = 0\end{math} (i.e., \begin{math}\lambda_{\mathrm{irr}} \neq 0 \ \mathrm{implies}\  \det[\rho_0]=0\ \mathrm{or}\  p^{Y_n,y_{n+1}}(c_{n+1})P[y_{n+1}|Y_n]  = 0\ \mathrm{for\ some\ }Y_n,y_{n+1},c_{n+1}\end{math}).
We can also prove that if
\begin{math}\det[\rho_0] \neq 0\  \mathrm{and}\ \det[M_{y_n}] \neq 0 \ \mathrm{for\ all\ }y_n,\ \mathrm{then}\ p^{Y_n,y_{n+1}}(c_{n+1})P[y_{n+1}|Y_n]  > 0\ \mathrm{for\ all} \ Y_n,y_{n+1},c_{n+1}
\end{math}
where the proof of this is given by
\begin{equation}
    \label{meas_fullrank_pr}
    \begin{split}
        &P[Y_n]p^{Y_n,y_{n+1}}(c_{n+1})P[y_{n+1}|Y_n]= \bra{c_{n+1}}P[Y_{n},y_{n+1}]\sigma_{t_n}^{Y_{n},y_{n+1}}\ket{c_{n+1}} \\
        &= \bra{c_{n+1}}\Bigl\{\sum_{z_{n+1},\psi_{t_n}|Y_{n}} \mathcal{M}_{y_{n+1},z_{n+1}}\mathcal{R}^{Y_{n}}_{d_{n}}\dots \mathcal{M}_{y_{1},z_{1}}p_0(a_0)\ket{a_0}\bra{a_0} \mathcal{M}_{y_1,z_{1}}^\dag \dots\mathcal{M}_{y_{n+1},z_{n+1}}^\dag\Bigr\} \ket{c_{n+1}}\\
        &\geq \sum_{a_0} \|\bra{c_{n+1}} \mathcal{M}_{y_{n+1},z_{n+1}=y_{n+1}}\mathcal{R}^{Y_{n}}_{d_{n}=0}\mathcal{M}_{y_{n},z_{n}=y_{n}}\dots \mathcal{R}^{Y_0}_{d_1=0}\mathcal{M}_{y_{1},z_{1}=y_{1}}\ket{a_0}\|^2 p_0(a_0)\\
        &= \bra{c_{n+1}} \mathcal{M}_{y_{n+1},y_{n+1}}\mathcal{R}^{Y_{n}}_{0}\dots\mathcal{R}^{Y_0}_{0}\mathcal{M}_{y_1,y_1}\rho_0 \mathcal{M}_{y_1,y_1}^\dag\mathcal{R}^{Y_0\dag}_{0}\dots\mathcal{R}^{Y_{n}\dag}_{0}\mathcal{M}_{y_{n+1},y_{n+1}}^\dag \ket{c_{n+1}} >0.\\
    \end{split}
\end{equation}
We used that \begin{math}\mathcal{M}_{y_{n+1},y_{n+1}}\mathcal{R}^{Y_{n}}_{0}\dots\mathcal{M}_{y_1,y_1}\rho_0 \mathcal{M}_{y_1,y_1}^\dag\dots\mathcal{R}^{Y_{n}\dag}_{0}\mathcal{M}_{y_{n+1},y_{n+1}}^\dag  \end{math} is full rank, positive definite, and Hermitian in order to show the inequality in the fourth line.
We therefore proved that \eqref{fullrank_cond} implies \begin{math}\lambda_{\mathrm{irr}} = 0 \end{math}.
Even though we do not show the necessary condition for \begin{math}\lambda_{\mathrm{irr}} =0\end{math} strictly, absolute irreversibility is expected to be nonzero (\begin{math}\lambda_{\mathrm{irr}} >0\end{math}), in most cases where the full rank condition \eqref{fullrank_cond} is broken.

\section{Fine unraveled dynamics}\label{Level fixing effect}
In this section, we explain the fine unraveled dynamics of the system in detail. In Section \ref{Time evolution of diagonal bases}, we show that diagonal bases $\{\ket{b_{n+1}}^{Y_n}\}, \{\ket{c_{n+1}}^{Y_{n+1}}\}$ of conditional density operators $\rho_{t_n}^{Y_n},\ \sigma_{t_n}^{Y_{n+1}}$ change smoothly as long as there is no measurement quantum jumps (i.e., $y_n=0$) and no pulse application (i.e., $V^{Y_n}=1$). In Section \ref{Time evolution of fine trajectory}, we illustrate the stochastic dynamics of the fine trajectory.

\subsection{Time evolution of diagonal bases}\label{Time evolution of diagonal bases}
Conditional density operators $\rho_{t_n}^{Y_n}$ and $\sigma_{t_n}^{Y_{n+1}}$ undergo smooth time evolution during no measurement quantum jump or sudden-pulse application. Therefore, their diagonal bases \begin{math}\ket{b_{n+1}}^{Y_n}\end{math} and \begin{math}\ket{c_{n+1}}^{Y_{n+1}}\end{math} are expected to change smoothly. In this subsection, we show this by using the perturbation theory. For a clear explanation, we assign numbers to diagonal bases as \begin{math}\{\ket{b_{n+1}=i}^{Y_n}\}_{i=1\dots \alpha},\{\ket{c_{n+1}=i}^{Y_{n+1}}\}_{i=1\dots \alpha}\end{math} so that the same number is assigned to smoothly shifted bases.

We can describe the no-jump time evolution of conditional density operators in \begin{math}[t_n,t_{n+1})\end{math} as
\begin{equation}
    \label{cond_ev_sigma}
    \begin{split}
        \sigma_{t_n}^{Y_n,0} &= \rho_{t_n}^{Y_n} + A\Delta t,\\
        A \equiv \Bigl\{\sum_{y} (1-\eta_{y}) \mathcal{D} [M_{y}]\rho_{t_n}^{Y_n} + \sum_{y}-\frac{\eta_{y}}{2}&\{M_{y}^\dag M_{y},\rho_{t_n}^{Y_n}\} + \eta_{y}\tr[M_{y}\rho_{t_n}^{Y_n}M_{y}^\dag]\rho_{t_n}^{Y_n}\Bigr\},\\
    \end{split}
\end{equation}
\begin{equation}
     \label{cond_ev_rho}
     \begin{split}
         \rho_{t_{n+1}}^{Y_n,0} &= \rho_{t_n}^{Y_n} + B\Delta t,\\
         B \equiv \Bigl\{-i[H^{Y_n}_{t_n}+h^{Y_n}_{t_n},\rho_{t_n}^{Y_n}] + \sum_{d}\mathcal{D}[L_{d}^{Y_n}]\rho_{t_n}^{Y_n} +\sum_{y} (1-\eta_{y}) &\mathcal{D} [M_{y}]\rho_{t_n}^{Y_n} + \sum_{y}-\frac{\eta_{y}}{2}\{M_{y}^\dag M_{y},\rho_{t_n}^{Y_n}\} + \eta_{y}\tr[M_{y}\rho_{t_n}^{Y_n}M_{y}^\dag]\rho_{t_n}^{Y_n}\Bigr\}.
     \end{split}
\end{equation}
We define the first order perturbation term of population $p^{Y_n}_{(1)} (b_{n+1})$ and diagonal basis \begin{math}\ket{b_{n+1}}^{Y_n}_{(1)}\end{math} as
\begin{equation}
\label{pop_dt}
    p^{Y_n,0}(c_{n+1}=i) = p^{Y_n}(b_{n+1} = i) + p^{Y_n}_{(1)} (b_{n+1}=i) \Delta t +\mathcal{O}(\Delta t^2),
\end{equation}
\begin{equation}
\label{ket_dt}
    \ket{c_{n+1}=i}^{Y_{n},0} = \ket{b_{n+1}=i}^{Y_n} + \ket{b_{n+1}=i}^{Y_n}_{(1)} \Delta t +\mathcal{O}(\Delta t^2).
\end{equation}
Then, the following equation holds:
\begin{equation}
    \label{eq_perturb}
    (p^{Y_n}(b_{n+1}) + p^{Y_n}_{(1)} (b_{n+1}) \Delta t+\dots)(\ket{b_{n+1}}^{Y_n} + \ket{b_{n+1}}^{Y_n}_{(1)} \Delta t +\dots) =(\rho_{t_n}^{Y_n} + A\Delta t)(\ket{b_{n+1}}^{Y_n} + \ket{b_{n+1}}^{Y_n}_{(1)} \Delta t +\dots) .
\end{equation}
We can determine \begin{math}p^{Y_n}_{(1)} (b_{n+1})\end{math} and \begin{math}\ket{b_{n+1}}^{Y_n}_{(1)}\end{math} from Eq.~\eqref{eq_perturb} as
\begin{equation}
    \label{P_perturb}
    p^{Y_n}_{(1)} (b_{n+1}=i) \equiv {}^{Y_n}\!\bra{b_{n+1}=i}A\ket{b_{n+1}=i}^{Y_n},
\end{equation}
\begin{equation}
    \label{ket_perturb}
    \ket{b_{n+1}=i}^{Y_n}_{(1)} \equiv \sum_{j\neq i} \frac{{}^{Y_n}\!\bra{b_{n+1}=j}A\ket{b_{n+1}=i}^{Y_n}}{p^{Y_n}(b_{n+1}=i)-p^{Y_n}(b_{n+1}=j)}\ket{b_{n+1}=j}^{Y_n}.
\end{equation}
In the same way, the eigenvalues and diagonal bases of \begin{math}\rho_{t_{n+1}}^{Y_n,0}\end{math} can be calculated as
\begin{equation}
    \label{pop_dt_rho}
    p^{Y_n,0}(b_{n+2}=i) =  p^{Y_n}(b_{n+1} = i) + {}^{Y_n}\!\bra{b_{n+1}=i}B\ket{b_{n+1}=i}^{Y_n} \Delta t +\mathcal{O}(\Delta t^2),
\end{equation}
\begin{equation}
    \label{ket_dt_rho}
    \ket{b_{n+2}=i}^{Y_n,0} = \ket{b_{n+1}=i}^{Y_n} + \sum_{j\neq i} \frac{{}^{Y_n}\!\bra{b_{n+1}=j}B\ket{b_{n+1}=i}^{Y_n}}{p^{Y_n}(b_{n+1}=i)-p^{Y_n}(b_{n+1}=j)}\ket{b_{n+1}=j}^{Y_n} \Delta t+ \mathcal{O}(\Delta t^2).
\end{equation}

From Eqs.~\eqref{ket_dt} and \eqref{ket_dt_rho}, it can be seen that the diagonal bases change continuously, except at the points where the populations of multiple bases are the same (i.e., $p^{Y_n}(b_{n+1}=i)=p^{Y_n}(b_{n+1}=j) \ \mathrm{for}\ i\neq j$) implying the degeneracies of \begin{math}\rho_{t_n}^{Y_n}\end{math}. The degeneracies can be avoided even with very small non-diagonal term, in which the bases are swapped smoothly (such phenomena is called avoided crossing). Therefore, the degeneracies of $\rho_{t_n}^{Y_n}$ are expected to be very rare events.

\subsection{Time evolution of the fine trajectory} \label{Time evolution of fine trajectory}
In this subsection, we rewrite the set of Kraus operators of the fine unraveling \eqref{fine_unraveling_eta} as
\begin{equation}
\{ \Pi_{b_{n+2}}^{Y_{n},y_{n+1}}\mathcal{R}_{d_n}^{Y_n} \Pi_{c_{n+1}}^{Y_{n},y_{n+1}} \mathcal{M}_{y_{n+1},z_{n+1}} \}_{d_{n+1},c_{n+1},y_{n+1},z_{n+1},b_{n+2}} ,
\end{equation} 
in order to focus on the time evolution of \begin{math}\ket{b_{n+1}}^{Y_n}\end{math}. If the quantum jump does not occur in $[t_n,t_{n+1})$, the transition probability of \begin{math}b_{n+1}\to c_{n+1}\end{math} and \begin{math}c_{n+1}\to b_{n+2}\end{math} can be calculated as 
\begin{equation}
\label{Zeno:meas}
    P[b_{n+1} =i\to c_{n+1} =j|Y_n,d_{n+1} =0,y_{n+1}=0,z_{n+1}=0] = 
    \begin{cases}
    1+\mathcal{O}(\Delta t^2) &(i =j)\\
    \mathcal{O}(\Delta t^2) &(i\neq j),\\
    \end{cases}
\end{equation}
\begin{equation}
\label{Zeno:diss}
    P[c_{n+1} =j \to b_{n+2}=k|Y_n,d_{n+1} =0,y_{n+1}=0,z_{n+1}=0] =
    \begin{cases}
    1+\mathcal{O}(\Delta t^2) &(j =k)\\
    \mathcal{O}(\Delta t^2) &(j \neq k).\\
    \end{cases}
\end{equation}
Here, we used Eqs.~\eqref{ket_dt} and \eqref{ket_dt_rho}, and that Kraus operators of no measurement $M_0$ and no dissipation $L_0$ can be represented as $M_0=1+\mathcal{O}(\Delta t)$ and $L_0=1+\mathcal{O}(\Delta t)$. 
Since the transition probabilities to different bases (i.e., $b_{n+1}\neq c_{n+1}$ or $c_{n+1}\neq b_{n+2}$) are \begin{math}\mathcal{O}(\Delta t^2)\end{math} if $d_{n+1}=0,z_{n+1}=0$, they vanish in the limit of \begin{math}\Delta t\to0\end{math}. 
Also, the probabilities that the measurement and dissipation quantum jumps occur in the same time step (i.e., $d_{n+1}\neq0 $ and $z_{n+1}\neq0$) are also \begin{math}\mathcal{O}(\Delta t^2)\end{math}, and thus negligible.
We note that Eq.~\eqref{Zeno:diss} holds even if the sudden pulse is applied at $t_{n+1}$, since the unitary gate $V^{Y_n}$ only rotates the diagonal bases and does not mix them.

From the above discussion, if the state at \begin{math}t_n\end{math} is \begin{math}\ket{b_{n+1}=i}^{Y_n}\end{math}, there are only the following four types of possible outcomes $(d_{n+1},c_{n+1},z_{n+1},y_{n+1},b_{n+2})$ in the next time step $[t_n,t_{n+1})$: (i) dissipation quantum jump occurs (i.e., $d_{n+1}\neq0,y_{n+1}=z_{n+1}=0$), (ii) measurement quantum jump occurs but is not detected (i.e., $d_{n+1}=0,y_{n+1}=0,z_{n+1}\neq0$), (iii) measurement quantum jump occurs and is detected (i.e., $d_{n+1}=0,y_{n+1}\neq0,z_{n+1}=y_{n+1}$), and (iv) no quantum jump occurs (i.e., $d_{n+1}=0,c_{n+1}=i,y_{n+1}=0,z_{n+1}=0,b_{n+2}=i$). 
Therefore, we can describe the fine unraveled dynamics of \begin{math}[t_n,t_{n+1})\end{math} in the form of the SME as
\begin{equation}
    \label{SME_b_t}
    \begin{split}
        \ket{b_{n+2}}^{Y_n,y_{n+1}} =\ket{b_{n+2}=i}^{Y_n,0} +\sum_{j}\ket{b_{n+2}=j}^{Y_n,0}&\Bigl\{\sum_{d_{n+1}} \Delta N_{b_{n+2}=j,d_{n+1}}^{Y_n} +\sum_{z_{n+1}\neq0} \Delta N_{b_{n+2}=j,z_{n+1},y_{n+1}=0}^{Y_n}\Bigr\}\\ &+\sum_{y_{n+1}\neq0,k}\ket{b_{n+2}=k}^{Y_n,y_{n+1}}\Delta N_{b_{n+2}=k,z_{n+1}=y_{n+1},y_{n+1}}^{Y_n},
    \end{split}
\end{equation}
if the state at \begin{math}t_n\end{math} is \begin{math}\ket{b_{n+1}=i}^{Y_n}\end{math}.
We here suppose that \begin{math}\rho_{t_n}^{Y_n}\end{math} is not degenerated.
The Poisson increments are defined as
\begin{equation}
    \label{poisson_level_fixing}
    \begin{split}
        \mathbb{E}[\Delta N_{b_{n+2}=j,d_{n+1}}^{Y_n}] &=\|{}^{Y_n,0}\!\bra{b_{n+2}=j}L_{d_{n+1}}^{Y_n}\ket{b_{n+1}=i}^{Y_n}\|^2 \Delta t,\\
        \mathbb{E}[\Delta N_{b_{n+2}=j,z_{n+1}\neq0,y_{n+1}=0}^{Y_n}] &=(1-\eta_{z_{n+1}})\|{}^{Y_n,0}\!\bra{b_{n+2}=j}M_{z_{n+1}}\ket{b_{n+1}=i}^{Y_n}\|^2 \Delta t,\\
        \mathbb{E}[\Delta N_{b_{n+2}=k,z_{n+1}=y_{n+1},y_{n+1}\neq0}^{Y_n}] &= \eta_{y_{n+1}}\|{}^{Y_n,y_{n+1}}\!\bra{b_{n+2}=k}M_{y_{n+1}}\ket{b_{n+1}=i}^{Y_n}\|^2 \Delta t.\\
    \end{split}
\end{equation}
Note that the system may not follow Eq.~\eqref{SME_b_t} at the point where the degeneracy of $\rho_{t_n}^{Y_n}$ occurs.

\section{Experiment-numerics hybrid verification method}
In this section, we give a detailed explanation on the experiment-numerics hybrid verification method of the generalized FT (Fig~\ref{fig:expnum_letter} of the main text). In Section \ref{Detailed explanation of the hybrid method}, we explain how and why we can verify the generalized FT under the fine trajectories by sampling the standard trajectories, and discuss the difference between the fine trajectory sampling and the hybrid verification method only with the standard trajectory sampling. In Section \ref{Numerical demonstration of the hybrid verification method}, we demonstrate the validity of the hybrid verification method through full numerical simulation by replacing the standard trajectory sampling with classical Monte-Carlo simulation.
\subsection{The hybrid method}\label{Detailed explanation of the hybrid method}
In the hybrid verification method, we evaluate the left-hand side of Eq.~\eqref{generalized_fluctuation_theorem} (i.e., $\langle e^{-\sigma -i_{\rm QC}}\rangle$) through the standard trajectory sampling. 
In order to evaluate $\langle e^{-\sigma -i_{\rm QC}}\rangle$, we introduce the exponential term $F[\psi_\tau]$ for each standard trajectory as
\begin{equation}
    \label{QC_trans_schroe}
         F[\psi_\tau]\equiv \sum_{\pi_\tau}\frac{P[\psi_\tau,\pi_\tau]}{P[\psi_\tau]} e^{-\sigma[\psi_\tau,\pi_\tau]-i_{\mathrm{QC}}[\psi_\tau,\pi_\tau]}.
\end{equation}
Under this definition, the ensemble average of $F[\psi_\tau]$ over the standard trajectories agrees with $\langle e^{-\sigma -i_{\rm QC}}\rangle$, which can be shown as
\begin{equation}
\label{expnum_mecha}
    \begin{split}
        \langle e^{-\sigma -i_{\rm QC}}\rangle &=\sum_{\psi_\tau,\pi_\tau} P[\psi_\tau,\pi_\tau]e^{-\sigma[\psi_\tau,\pi_\tau]-i_{\mathrm{QC}}[\psi_\tau,\pi_\tau]}\\
        &=\sum_{\psi_\tau}P[\psi_\tau]F[\psi_\tau].
    \end{split}
\end{equation}
Therefore, the left-hand side of Eq.~\eqref{generalized_fluctuation_theorem} can be evaluated by sampling the standard trajectories $\psi_\tau$, calculating $F[\psi_\tau]$ for each trajectory, and taking the average of $F[\psi_\tau]$.

We here make a remark on Eq.~\eqref{expnum_mecha}. Although there is the correspondence between the bundle of the fine trajectories $\{\psi_\tau,\pi_\tau\}_{\pi_\tau}$ and single standard trajectory $\psi_\tau$, their realization probabilities are in general not the same:
\begin{equation}
    \label{exp_verify_P_fuicchi}
    P[\psi_\tau] \neq \sum_{\pi_\tau}P[\psi_\tau,\pi_\tau].
\end{equation}
Therefore, the convergence rates of the ensemble average of $F[\psi_\tau]$ over the standard trajectories and that of $e^{-\sigma[\psi_\tau,\pi_\tau]-i_{\mathrm{QC}}[\psi_\tau,\pi_\tau]}$ over the fine trajectories are not the same. Namely, required sampling number may both increase or decrease. Furthermore, even the converged values might not coincide if the contribution of $(\psi_\tau,\pi_\tau)$ satisfying $P[\psi_\tau]=0$ but $P[\psi_\tau,\pi_\tau] \neq 0$ is not negligible. However, we  expect that this subtle problem can be avoided in practice, as numerically verified in Section \ref{Numerical demonstration of the hybrid verification method}.

We here explain that $F[\psi_\tau]$ for each trajectory can be exactly calculated with reasonable numerical cost.
The important point is that the number of the realizable fine trajectories with \begin{math}P[\psi_\tau,\pi_\tau] \neq 0\end{math} corresponding to single standard trajectory $\psi_\tau$ is finite even in the continuous time limit \begin{math}\Delta t\to 0\end{math}. This is because under the fine unraveling, transition between the different diagonal bases (i.e., $b_{n+1}\neq c_{n+1}$ or $c_{n+1}\neq b_{n+2}$) does not occur while there is no quantum jump, as explained in Section \ref{Time evolution of fine trajectory}. Therefore, the number of the realizable fine trajectories corresponding to $\psi_\tau$ is given by \begin{math}\alpha^{J[\psi_\tau]}\end{math}, where \begin{math}J[\psi_\tau]\end{math} denotes the total number of quantum jumps in the trajectory \begin{math}\psi_\tau\end{math} and $\alpha$ represents the Hilbert space dimension of the system, as defined above. Equation~\eqref{QC_trans_schroe} can be calculated exactly in a classical computer since the infinite sum taken in Eq.~\eqref{QC_trans_schroe} is in fact only the sum of \begin{math}\alpha^{J[\psi_\tau]}\end{math} trajectories, ignoring the terms whose values are 0.

\subsection{Numerical demonstration of the hybrid method} \label{Numerical demonstration of the hybrid verification method}
\begin{figure}[tb]
\begin{center}
\includegraphics[width=0.4\textwidth]{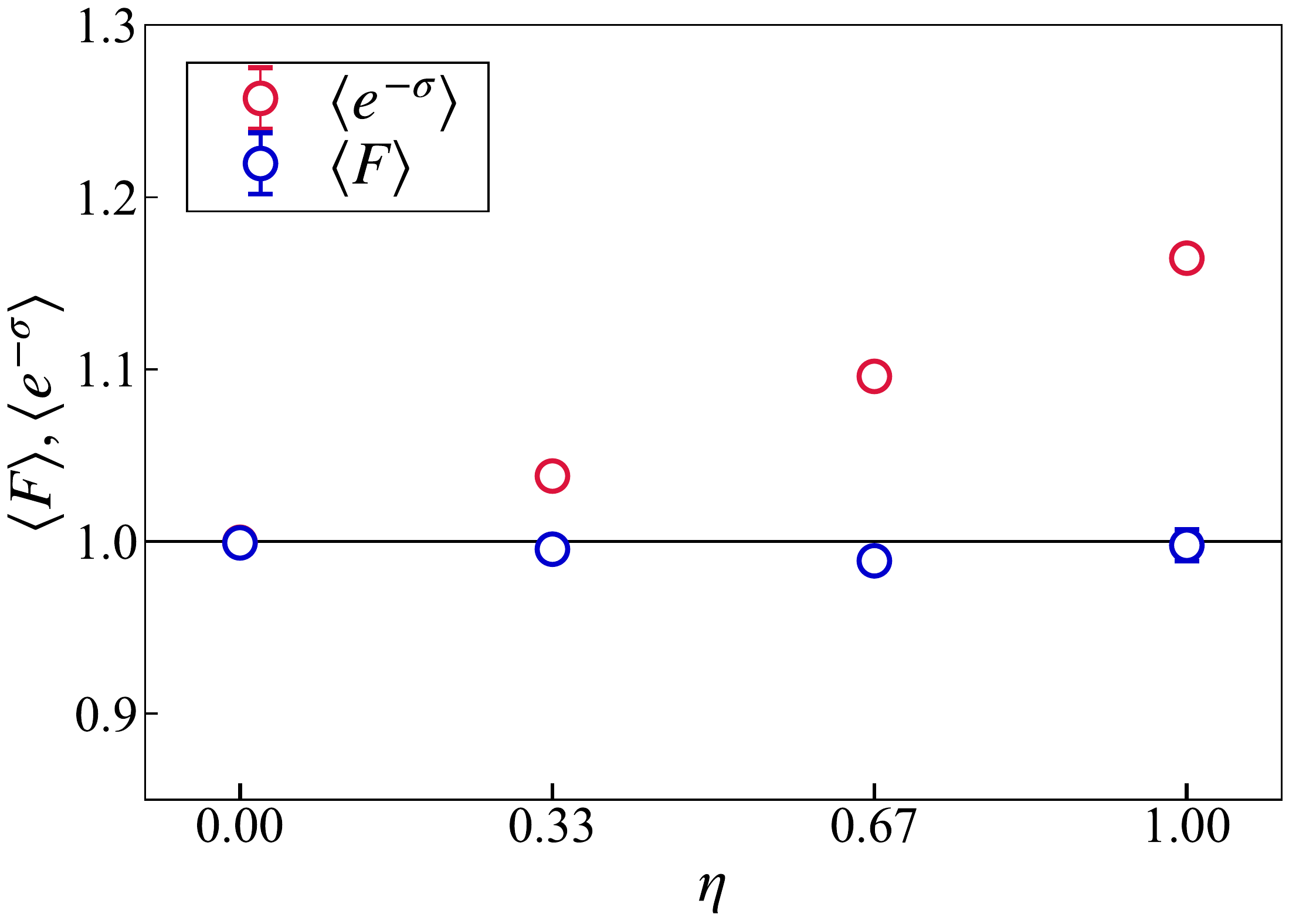}
\caption{Numerical results by the standard trajectory sampling. In this numerical simulation, we replace the experimental sampling of the standard trajectories with classical Monte-Carlo simulation. The average values of \begin{math}e^{-\sigma[\psi_\tau]}\end{math} and \begin{math}F[\psi_\tau]\end{math} are plotted. The system parameters are taken to be identical to those in Fig.~\ref{fig:fluctuation}. We fix the trial time to $\tau=10$ and change the detection rate as \begin{math}\eta=0.00,0.33,0.67,1.00\end{math}. \begin{math}1.0\times 10^4\end{math} standard trajectories are sampled for each data point.}
\label{fig:num_verif_Sch}
\end{center}
\end{figure}
In this subsection, we numerically demonstrate the hybrid method, by replacing the experimental sampling part of the hybrid method with classical Monte-Carlo simulation on a classical computer. Since the probability distributions of the standard trajectories and fine trajectories are different (Eq.~\eqref{exp_verify_P_fuicchi}), numerical calculation performed in this subsection is different from the simulation in Section \ref{Numerical calculation}, in which the fine unraveled dynamics is calculated. 
Figure \ref{fig:num_verif_Sch} shows the results of the simulation. From Fig.~\ref{fig:num_verif_Sch}, we can see that the average value of $\langle F\rangle$ coincides with 1 under arbitrary detection rate \begin{math}\eta\end{math}, and thus the generalized FT is verified. This result shows that the average value  \begin{math}F[\psi_\tau]\end{math} in the hybrid method converges to the expected value if we sample sufficient number of trajectories, and shows the validity of the experiment-numerics hybrid verification method.

\section{Comparison to the classical stochastic process}
In this section, we first introduce the classical stochastic process corresponding to the setup discussed in this work, where we refer to the former setup as the classical setup, and the latter as the quantum setup. 
Then, we explain that the generalized FTs derived in the quantum setup in this work, can be interpreted as the quantum counterpart of the generalized FTs derived in the classical setup in the previous works \cite{sagawa2012nonequilibrium,ito2013information,ito2016information}.

\begin{figure}[tb]
\begin{center}
\includegraphics[width=0.4\textwidth]{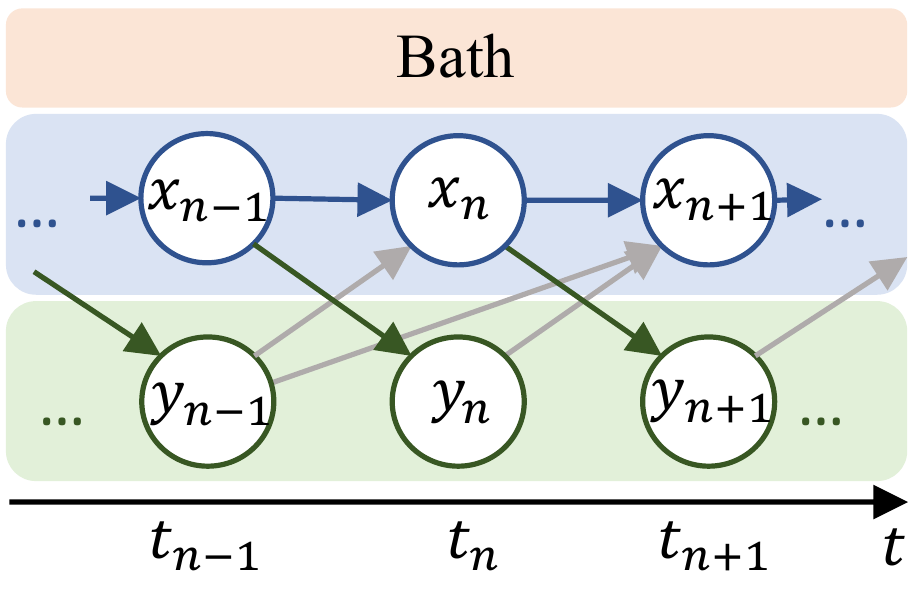}
\caption{The causal relationship of the classical stochastic process corresponding to the setup in this work (Fig.~\ref{fig:setup}(b) of the main text).}
\label{fig:classical_case}
\end{center}
\end{figure}
First, we introduce the classical setup. 
We consider a classical system interacting with a heat bath at inverse temperature $\beta$ and continuously measured. 
Here, we descritize time as $t_n=n\Delta t$, consider the time evolution from $t=0$ to $t=\tau \equiv t_N$ and later take the continuous time limit $\Delta t\to 0$.
We define \begin{math}x_n\end{math} as the system's state at $t_n$, and \begin{math}X_n = (x_1,x_2,\dots,x_{n})\end{math} as the trajectory of the system's state until \begin{math}t_n\end{math}. We also define \begin{math}y_n\end{math} as the measurement outcome obtained at $t=t_n$, and $Y_n \equiv (y_1,y_2,\dots y_n)$ represents all the measurement outcomes until $t=t_n$. 
Figure \ref{fig:classical_case} shows the causal relationship of the classical setup corresponding to the quantum setup (Fig.~\ref{fig:setup}(b) of the main text). We here assume that $y_{n+1}$ is completely determined by the system state at $t_n$ (i.e., \begin{math}P[y_{n+1}|X_{n},Y_n] = P[y_{n+1}|x_{n}]\end{math}). Continuous feedback is performed by changing the control protocol of the system in $[t_n,t_{n+1})$ depending on the measurement outcomes until $t_n$ (i.e., $Y_n$). Note that the control protocol in $[t_n,t_{n+1})$ does not depend on the measurement results obtained after $t_{n+1}$ (i.e., $(y_{n+1},\dots,y_N)$), because of causality.

According to the previous works, the generalized FTs in this classical setup can be described as
\begin{equation}
    \label{generalized FT:classcont_SU}
    \begin{split}
       \langle e^{-\sigma-i_{\mathrm{TE}}} \rangle &=1,\\
       \langle i_{\mathrm{TE}}\rangle &= \sum_{n=0}^{N-1} I(x_n:y_{n+1}|Y_n),
    \end{split}
\end{equation}
and
\begin{equation}
    \label{generalized FT:classcont_IS}
    \begin{split}
        \langle e^{-\sigma-\Theta} \rangle &=1,\\
        \langle \Theta \rangle &= \langle i_{\mathrm{TE}}\rangle -I(x_N:Y_N),
    \end{split}
\end{equation}
where \begin{math}I(x_n:y_{n+1}|Y_n)\equiv \sum_{Y_n}P[Y_n](\mathcal{H}_{Y_n}(x_n)-\sum_{y_n}P[y_{n+1}|Y_n]\mathcal{H}_{Y_n,y_{n+1}}(x_n))\end{math} denotes conditional mutual information, and its accumulation \begin{math}\langle i_{\mathrm{TE}}\rangle \end{math} is called transfer entropy. As mentioned in the main text, the transfer entropy corresponds to the QC-transfer entropy in that their increment in \begin{math}[t_n,t_{n+1})\end{math} is the information transfer under the given measurement outcomes $Y_n$.

We next discuss the correspondence of the generalized FTs in the classical and quantum setups.
From the correspondece of the transfer entropy and the QC-transfer entropy, we can see that Eq.~\eqref{generalized FT:classcont_SU} corresponds to our main result Eq.~\eqref{generalized_fluctuation_theorem} of the main text. In fact, we can also derive another type of generalized FT in the quantum setup, which is the quantum counterpart of Eq.~\eqref{generalized FT:classcont_IS}:
\begin{equation}
    \label{generalized FT:QC_ito}
    \begin{split}
        \langle e^{-\sigma-\Theta_{\mathrm{QC}}} \rangle &=1,\\
        \langle \Theta_{\mathrm{QC}} \rangle &= \langle i_{\mathrm{QC}}\rangle - \{S(\rho_\tau) -\sum_{Y_N}P[Y_N]S(\rho_\tau^{Y_N}) \}.
    \end{split}
\end{equation}
The unraveling used for the derivation of Eq.~\eqref{generalized FT:QC_ito} is almost the same as the fine unraveling. Only the difference is that we insert the additional PM in the diagonalized basis of \begin{math}\rho_\tau^{Y_N} \equiv \sum_{b} p^{Y_N}(b)\ket{b}^{Y_N}{}^{Y_N\!\!}\bra{b}\end{math}, right before the final PM in the diagonalized basis of \begin{math}\rho_{\tau}\end{math}. The outcome for the inserted PM is defined as $b_{N+1}$, and therefore, the trajectory is designated by $\{\psi_\tau,\pi_\tau,b_{N+1}\}$ in this unraveling.
We can derive Eq.~\eqref{generalized FT:QC_ito} by defining the stochastic information term $\Theta_{\rm QC}$ as \begin{math}\Theta_{\mathrm{QC}}[\psi_\tau,\pi_\tau,b_{N+1}] \equiv i_{\mathrm{QC}}[\psi_\tau,\pi_\tau] -\ln{p^{Y_N}(b_{N+1})} +\ln{p_\tau(a_\tau)}\end{math}.
We can derive the generalized SL from Eqs.~\eqref{generalized_fluctuation_theorem}, \eqref{generalized FT:classcont_SU}, \eqref{generalized FT:classcont_IS} and \eqref{generalized FT:QC_ito} as  \begin{math} \langle\sigma\rangle \geq-\langle \Theta\rangle \geq -\langle i\rangle\end{math} and \begin{math}\langle\sigma\rangle \geq-\langle \Theta_{\rm QC}\rangle \geq -\langle i_{\rm QC}\rangle\end{math}. Here, Eqs.~\eqref{generalized FT:classcont_IS} and \eqref{generalized FT:QC_ito} give tighter bounds of the entropy production than Eqs.~\eqref{generalized FT:classcont_SU} and \eqref{generalized_fluctuation_theorem}, respectively.

\section{generalized fluctuation theorem under the standard unraveling}
As mentioned in the main text, the generalized FT can be derived under the standard unraveling, in use of the correction term \begin{math}i_{\mathrm{QJT}}\end{math}. In this section, we provide the proof of the generalized FT for the standard trajectories. In Section \ref{generalized FT in Schrodinger unraveling}, we introduce the QJT information, which is incorporated in the generalized FT under the standard trajectory. In the second subsection, we give the proof of the generalized FT. We note that the perfect measurement (i.e., \begin{math}\eta_{y} =1\end{math}) is supposed in the derivation in this section, and therefore the subscript \begin{math}z_n\end{math} is omitted.
\subsection{Introduction of the QJT information} \label{generalized FT in Schrodinger unraveling}
The generalized FT under the standard unraveling is derived as 
\begin{equation}
    \label{generalized FTSch}
    \langle e^{-\sigma-i_{\mathrm{QJT}}} \rangle =1-\lambda_{\mathrm{irr}}.
\end{equation}
The correction term $i_{\mathrm{QJT}}$ is defined as 
\begin{equation}
    \label{QJTinfo_ren}
    \begin{split}
        i_{\mathrm{QJT}}[\psi_\tau] &\equiv \sum_{n=0}^{N-1} \ln{\|\mathcal{M}^\dag_{y_{n+1}}\mathcal{N}[\mathcal{R}^{Y_{n}\dag}_{d_{n+1}}\ket{\bar{\psi}(t_{n+1},\tau)}]\|^2} -\ln{P[y_{n+1}|Y_{n}]}, \\
        \langle i_{\mathrm{QJT}}\rangle &=\sum_{n=0}^{N-1} \sum_{Y_{n}}P[Y_{n}] \Big\{\sum_{y_{n+1}}P[y_{n+1}|Y_n] \mathcal{I}_C(\sigma_{t_n}^{Y_n,y_{n+1}}:\{\mathcal{L}^{Y_n,y_{n+1}}_{\psi_{(t_{n+1},\tau)}} \mathcal{R}^{Y_{n}}_{d_{n+1}}\}_{\psi_{(t_{n+1},\tau)},d_{n+1}}) -\mathcal{I}_C(\rho_{t_n}^{Y_n}:\{\mathcal{L}^{Y_n}_{\psi_{(t_n,\tau)}}\}_{\psi_{(t_n,\tau)}}) \Big\},
    \end{split}
\end{equation}
where \begin{math}\ket{\bar{\psi}}\end{math} and \begin{math}\mathcal{L}\end{math} are defined as
\begin{align}
    \label{invsch_stateevl}
    \begin{split}
        &\ket{\bar{\psi}(t_n,\tau)} \equiv \mathcal{N}[\mathcal{L}^{Y_N\dag}_{\psi_{(t_n,\tau)}} \ket{a_\tau}], \\
        \mathcal{L}^{Y_N}_{\psi_{(t_n,\tau)}} &\equiv \Pi_{a_\tau}^\tau \mathcal{L}^{Y_N}_{d_{N},y_{N}} \dots \mathcal{L}^{Y_n}_{d_{n+1},y_{n+1}}.\\
    \end{split}
\end{align}
Here, \begin{math}\ket{\bar{\psi}(t_n,\tau)}\end{math} represents the state of the inverse process (as will be explained in detail in Section \ref{Derivation of generalized FT in Schrodinger unraveling}).
We introduce \begin{math}\psi_{(t_n,\tau)}\end{math} to designate the standard trajectory in \begin{math}[t_n,\tau]\end{math} (i.e., $\psi_{(t_n,\tau)} \equiv ((d_{n+1},d_{n+2}\dots,d_{N}),(y_{n+1},\dots,y_{N}),a_\tau)$) and \begin{math}\mathcal{L}^{Y_N}_{\psi_{(t_{n},\tau)}}\end{math} represents the corresponding Kraus operator.
We define \begin{math}\mathcal{I}_C(\rho:\{M_x\}_x) \equiv H(p_{\rho}^{M_x} \|p_{\frac{1}{d}}^{M_x})\end{math} as the classical relative entropy between \begin{math}p_{\rho}^{M_x}\end{math} and \begin{math}p_{\frac{1}{d}}^{M_x}\end{math}. Here, \begin{math}p_{\rho}^{M_x}\end{math} denotes the probability distribution of the outcome of the measurement $\{M_x\}_x$ on \begin{math}\rho\end{math}. This quantity measures how much we can specify the quantum state based on the measurement results \cite{gong2016quantum}. 

The increment of \begin{math}\langle i_{\mathrm{QJT}}\rangle\end{math} in \begin{math}[t_n,t_{n+1})\end{math} (i.e., \begin{math}P[y_{n+1}|Y_n] \mathcal{I}_C(\sigma_{t_n}^{Y_n,y_{n+1}}:\{\mathcal{L}^{Y_n,y_{n+1}}_{\psi_{(t_{n+1},\tau)}} \mathcal{R}^{Y_{n}}_{d_{n+1}}\}) -\mathcal{I}_C(\rho_{t_n}^{Y_n}:\{\mathcal{L}^{Y_n}_{\psi_{(t_{n},\tau)}}\})\end{math}) is represented as the QJT information \begin{math}\mathcal{I}_{\mathrm{QJT}}\end{math} introduced in Ref.~\cite{gong2016quantum}. Therefore, we name this correction term $i_{\rm QJT}$ \emph{total QJT information}. QJT information represents the difference of knowledge on pre-measurement state \begin{math}\rho_{t_n}^{Y_n}\end{math} and post-measurement state \begin{math}\sigma_{t_{n+1}}^{Y_n,y_{n+1}}\end{math}, obtained by the selective measurement outcomes in $[t_n,\tau]$ (i.e., \begin{math}\psi_{(t_n,\tau)}\end{math}). It is obvious that \begin{math}\langle i_{\mathrm{QJT}}\rangle\end{math} and \begin{math}\mathcal{I}_{\mathrm{QJT}}\end{math} depend on how the system-bath interaction is unraveled by the Lindblad operators, and does not uniquely determined by the SME \eqref{SME_eta}.

The ensemble average of total QJT information \begin{math}\langle i_{\mathrm{QJT}}\rangle\end{math} is calculated as follows:
\begin{equation}
    \label{QJT_ens_calc}
    \begin{split}
        \langle i_{\mathrm{QJT}}\rangle &= \sum_{\psi_\tau} P[\psi_\tau]\sum_{n=0}^{N-1} \ln{\|\mathcal{M}^\dag_{y_{n+1}}\mathcal{N}[\mathcal{R}^{Y_{n}\dag}_{d_{n+1}}\ket{\bar{\psi}(t_{n+1},\tau)}]\|^2} -\ln{P[y_{n+1}|Y_{n}]}\\
        &= \sum_{n=0}^{N-1}\sum_{\psi_\tau}P[\psi_\tau] 
        \Biggl\{ \ln{\frac{\tr[\mathcal{L}_{d_{n+1},y_{n+1}}^{Y_n\dag}\mathcal{L}_{d_{n+2},y_{n+2}}^{Y_{n+1}\dag} \dots \mathcal{L}_{d_{N},y_{N}}^{Y_{N-1}\dag} \ket{a_\tau}\bra{a_\tau}\mathcal{L}_{d_{N},y_{N}}^{Y_{N-1}}\dots \mathcal{L}_{d_{n+2},y_{n+2}}^{Y_{n+1}}\mathcal{L}_{d_{n+1},y_{n+1}}^{Y_n}]}
        {\tr[\mathcal{R}^{Y_n\dag}_{d_{n+1}}\mathcal{L}_{d_{n+2},y_{n+2}}^{Y_{n+1}\dag} \dots \mathcal{L}_{d_{N},y_{N}}^{Y_{N-1}\dag} \ket{a_\tau}\bra{a_\tau}\mathcal{L}_{d_{N},y_{N}}^{Y_{N-1}}\dots \mathcal{L}_{d_{n+2},y_{n+2}}^{Y_{n+1}}\mathcal{R}^{Y_n}_{d_{n+1}}]}} \\
         &\ \ \ \ \ \ \ \ \ \ \ \ \ \ \ \ \ \ \ \ \ \ \ \ \ \ \ \ \ \ \ \ \ \ \ \ 
         -\ln{\frac{\bra{a_\tau}\mathcal{L}_{d_{N},y_{N}}^{Y_{N-1}} \dots\mathcal{R}^{Y_n}_{d_{n+1}}\mathcal{M}_{y_{n+1}} \rho_{t_n}^{Y_n} \mathcal{M}_{y_{n+1}}^\dag \mathcal{R}^{Y_n\dag}_{d_{n+1}} \dots \mathcal{L}_{d_{N},y_{N}}^{Y_{N-1}\dag}\ket{a_\tau} }
         {\bra{a_\tau}\mathcal{L}_{d_{N},y_{N}}^{Y_{N-1}}\dots \mathcal{R}^{Y_n}_{d_{n+1}} \sigma_{t_n}^{Y_n,y_{n+1}}  \mathcal{R}^{Y_n\dag}_{d_{n+1}}\dots\mathcal{L}_{d_{N},y_{N}}^{Y_{N-1}\dag}\ket{a_\tau}}}\Biggr\} \\
         &= \sum_{n=0}^{N-1}\Biggl\{\sum_{\psi_\tau}P[\psi_\tau] 
         \ln{\frac{\bra{a_\tau}\mathcal{L}_{d_{N},y_{N}}^{Y_{N-1}}\dots \mathcal{R}^{Y_n}_{d_{n+1}} \sigma_{t_n}^{Y_n,y_{n+1}}  \mathcal{R}^{Y_n\dag}_{d_{n+1}}\dots\mathcal{L}_{d_{N},y_{N}}^{Y_{N-1}\dag}\ket{a_\tau}}{\bra{a_\tau}\mathcal{L}_{d_{N},y_{N}}^{Y_{N-1}}\dots \mathcal{R}^{Y_n}_{d_{n+1}} \frac{1}{d} \mathcal{R}^{Y_n\dag}_{d_{n+1}}\dots\mathcal{L}_{d_{N},y_{N}}^{Y_{N-1}\dag}\ket{a_\tau}}}\\
         &\ \ \ \ \ \ \ \ -\sum_{\psi_\tau}P[\psi_\tau] 
         \ln{\frac{\bra{a_\tau}\mathcal{L}_{d_{N},y_{N}}^{Y_{N-1}} \dots\mathcal{L}_{d_{n+1},y_{n+1}}^{Y_{n}} \rho_{t_n}^{Y_n} \mathcal{L}_{d_{n+1},y_{n+1}}^{Y_{n}\dag}\dots \mathcal{L}_{d_{N},y_{N}}^{Y_{N-1}\dag}\ket{a_\tau}}{\bra{a_\tau}\mathcal{L}_{d_{N},y_{N}}^{Y_{N-1}} \dots\mathcal{L}_{d_{n+1},y_{n+1}}^{Y_{n}} \frac{1}{d} \mathcal{L}_{d_{n+1},y_{n+1}}^{Y_{n}\dag} \dots \mathcal{L}_{d_{N},y_{N}}^{Y_{N-1}\dag}\ket{a_\tau}}}\Biggr\}\\
          &= \sum_{n=0}^{N-1} \sum_{Y_n,y_{n+1}} P[Y_n,y_{n+1}]H(p_{\sigma_{t_n}^{Y_n,y_{n+1}}}^{\mathcal{L}^{Y_n,y_{n+1}}_{\psi(t_{n+1},\tau)}\mathcal{R}^{Y_n}_{d_{n+1}}} \|p_{\frac{1}{d}}^{\mathcal{L}^{Y_n,y_{n+1}}_{\psi(t_{n+1},\tau)}\mathcal{R}^{Y_n}_{d_{n+1}}}) 
          -\sum_{Y_n} P[Y_n]H(p_{\rho_{t_n}^{Y_n}}^{\mathcal{L}^{Y_n}_{\psi(t_n,\tau)}} \|p_{\frac{1}{d}}^{\mathcal{L}^{Y_n}_{\psi(t_n,\tau)}}) \\
          &= \sum_{n=0}^{N-1} \sum_{Y_n} P[Y_n]\big\{ \sum_{y_{n+1}} P[y_{n+1}|Y_n] \mathcal{I}_C (\sigma_{t_n}^{Y_n,y_{n+1}}:\{\mathcal{L}^{Y_n,y_{n+1}}_{\psi(t_{n+1},\tau)}\mathcal{R}^{Y_n}_{d_{n+1}}\}_{\psi_{(t_{n+1},\tau)},d_{n+1}}) -\mathcal{I}_C (\rho_{t_n}^{Y_n}:\{\mathcal{L}^{Y_n}_{\psi(t_n,\tau)}\}_{\psi_{(t_n,\tau)}})\big\}.
    \end{split}
\end{equation}
\subsection{Derivation of the the generalized FT under the standard unraveling} \label{Derivation of generalized FT in Schrodinger unraveling}
In this subsection, we derive the generalized FT under the standard unraveling.
The outline of the derivation is the same as that of the fine unraveling explained in Section \ref{Overview of the derivation}.
The probability of the inverse trajectory corresponding to the forward standard trajectory \begin{math}\psi_\tau\end{math} is defined as
\begin{equation}
    \label{invsch}
    \bar{P}[\psi_\tau] \equiv P[Y_N] \| \langle a_0|\bar{\phi}(0,\tau)\rangle\|^2\|\bar{\mathcal{R}}^{Y_N}_{d_{1}}\ket{\bar{\phi}(t_1,\tau)}\|^2 \dots \|\bar{\mathcal{R}}^{Y_{N}}_{d_{N-1}}\ket{\bar{\phi}(t_{N-1},\tau)}\|^2 \|\bar{\mathcal{R}}^{Y_{N}}_{d_{N}}\ket{a_\tau}\|^2 p_\tau(a_\tau),
\end{equation}
where \begin{math}\ket{\bar{\phi}}\end{math} and \begin{math}\bar{\mathcal{R}}\end{math} are defined as 
\begin{equation}
\label{Sch_inv_joutai}
\ket{\bar{\phi}(t_n,\tau)}\equiv
    \begin{cases}
    \ket{\bar{\psi}(t_n,\tau)} &(\ket{\bar{\psi}(t_m,\tau)} \neq 0 \mathrm{\ for\ all}\ m\ \mathrm{s.t.}\ m>n)\\
    \ket{\phi_0} \ &(\mathrm{otherwise}),\\
    \end{cases}
\end{equation}
\begin{equation}
    \label{Sch_inv_Kraus}
    \bar{\mathcal{R}}^{Y_{N}}_{d_{n+1}} \equiv 
    \begin{cases}
    L_0^{Y_n\dag}V^{Y_n\dag} &(d_{n+1} =0)\\
    L_{d^\prime}^{Y_n}\sqrt{\Delta t}V^{Y_n\dag} &(d_{n+1} = d).\\
    \end{cases}
\end{equation}
Here, $L_{d^\prime}^{Y_n}$ satisfies the detailed balance condition $L_{d^\prime}^{Y_n}=L_{d_{n+1}}^{Y_n\dag}e^{-\beta\frac{\Delta_{d_{n+1}}}{2}}$, and $\ket{\phi_0}$ is an arbitrary fixed pure state.
We note that if the forward trajectory is realizable (i.e., \begin{math}\psi\in\mathcal{V}\end{math}), \begin{math}\ket{\bar{\psi}(t_n,\tau)} \neq 0\end{math} always holds, and therefore \begin{math}\ket{\bar{\phi}(t_n,\tau)}=\ket{\bar{\psi}(t_n,\tau)}\end{math} holds for its corresponding inverse trajectory. Therefore, the proof of the generalized FT does not depend on the definition of $\ket{\phi_0}$.
We can show that this inverse process satisfies the probability normalization condition \begin{math}\sum_{\psi_\tau}\bar{P}[\psi_\tau] =1\end{math} as 
\begin{equation}
    \label{Sch_normal_cond}
    \begin{split}
        \sum_{\psi_\tau}\bar{P}[\psi_\tau]&=\sum_{Y_N}\sum_{\psi_\tau|Y_N}P[Y_N] \| \langle a_0|\bar{\phi}(0,\tau)\rangle\|^2\|\bar{\mathcal{R}}^{Y_N}_{d_{1}}\ket{\bar{\phi}(t_1,\tau)}\|^2 \dots \|\bar{\mathcal{R}}^{Y_{N}}_{d_{N}}\ket{a_\tau}\|^2 p_\tau(a_\tau)\\
        &=\sum_{Y_N}P[Y_N] \sum_{a_\tau}p_\tau(a_\tau)\Biggl[\sum_{d_{N}} \|\bar{\mathcal{R}}^{Y_{N}}_{d_{N}}\ket{a_\tau}\|^2
        \Bigl\{\sum_{d_{N-1}}\|\bar{\mathcal{R}}^{Y_{N}}_{d_{N-1}}\ket{\bar{\phi}(t_{N-1},\tau)}\|^2 \dots\\
        &\ \ \ \ \ \ \ \ \ \ \ \ \ \ \ \ \ \ \ \ \ \ \ \ \ \ \ \ \ \ \ \ \ \ \ \ \dots\bigl( \sum_{d_1}\|\bar{\mathcal{R}}^{Y_N}_{d_{1}}\ket{\bar{\phi}(t_1,\tau)}\|^2(\sum_{a_0} \|\langle a_0|\bar{\phi}(0,\tau)\rangle\|^2 )\bigr)\dots \Bigr\}\Biggr] \\
        &= \sum_{Y_N}P[Y_N]\  1 = 1 .\\
    \end{split}
\end{equation}
From the second to the third line, we used that \begin{math}\{\bar{\mathcal{R}}^{Y_N}_{d_{n}}\}\end{math} satisfies the completeness condition.

By using this inverse process, we can derive the GDFT as 
\begin{equation}
    \label{generalized FTsch_derivation}
    \begin{split}
     \frac{\bar{P}[\psi_\tau]}{P[\psi_\tau]} &= \frac{P[Y_N]p_\tau(a_\tau) \| \langle a_0|\bar{\phi}(0,\tau)\rangle\|^2\|\bar{\mathcal{R}}^{Y_N}_{d_{1}}\ket{\bar{\phi}(t_1,\tau)}\|^2 \dots \|\bar{\mathcal{R}}^{Y_{N}}_{d_{N-1}}\ket{\bar{\phi}(t_{N-1},\tau)}\|^2 \|\bar{\mathcal{R}}^{Y_{N}}_{d_{N}}\ket{a_\tau}\|^2 }
    {p_0(a_0)\|\bra{a_\tau}\mathcal{L}^{Y_N}_{\psi_{(0,\tau)}} \ket{a_0}\|^2}\\
    &= e^{-\Delta S}\frac{P[Y_N] \| \langle a_0|\bar{\phi}(0,\tau)\rangle\|^2\|\bar{\mathcal{R}}^{Y_N}_{d_{1}}\ket{\bar{\phi}(t_1,\tau)}\|^2 \dots \|\bar{\mathcal{R}}^{Y_{N}}_{d_{N-1}}\ket{\bar{\phi}(t_{N-1},\tau)}\|^2 \|\bar{\mathcal{R}}^{Y_{N}}_{d_{N}}\ket{a_\tau}\|^2}
    {\|\bra{a_0}\mathcal{M}_{y_1}^{\dag}\mathcal{R}_{d_1}^{Y_0\dag} \mathcal{M}_{y_{2}}^\dag\mathcal{R}_{d_{2}}^{Y_{1}\dag}\dots \mathcal{M}_{y_{N}}^\dag\mathcal{R}_{d_{N}}^{Y_{N-1}\dag}\ket{a_\tau}\|^2}\\
    &= P[Y_N] e^{-\Delta S} \frac{\|\bar{\mathcal{R}}^{Y_N}_{d_{1}}\ket{\bar{\psi}(t_1,\tau)}\|^2}{\|\mathcal{M}_{y_{1}}^\dag\mathcal{R}^{Y_0\dag}_{d_{1}}\ket{\bar{\psi}(t_1,\tau)}\|^2} \dots 
    \frac{\|\bar{\mathcal{R}}^{Y_{N}}_{d_{N-1}}\ket{\bar{\psi}(t_{N-1},\tau)}\|^2}{\|\mathcal{M}_{y_{N-1}}^\dag\mathcal{R}_{d_{N-1}}^{Y_{N-2}\dag}\ket{\bar{\psi}(t_{N-1},\tau)}\|^2}
    \frac{\|\bar{\mathcal{R}}^{Y_{N}}_{d_{N}}\ket{a_\tau}\|^2}{\|\mathcal{M}_{y_{N}}^\dag\mathcal{R}_{d_{N}}^{Y_{N-1}\dag}\ket{a_\tau}\|^2}\\
    &= e^{-\sigma}\prod_{n=0}^{N-1} \frac{P[y_{n+1}|Y_n]}{\|\mathcal{M}_{y_{n+1}}^\dag\mathcal{N}[\mathcal{R}^{Y_{n}\dag}_{d_{n+1}}\ket{\bar{\psi}(t_{n+1},\tau)}]\|^2} \\
    &= e^{-\sigma -i_{\mathrm{QJT}}}. \\
    \end{split}
\end{equation}

Finally, we show a sufficient condition for \begin{math}\lambda_{\mathrm{irr}}=0\end{math}. By comparing \begin{math}P[\psi_\tau]\end{math} and \begin{math}\bar{P}[\psi_\tau]\end{math}, we can see which trajectories are in the set \begin{math}\mathcal{V}\smallsetminus\mathcal{W}\end{math} that is the origin of \begin{math}\lambda_{\mathrm{irr}}\end{math}. The trajectories in \begin{math}\mathcal{V}\smallsetminus\mathcal{W}\end{math} satisfy $\|\mathcal{R}^{Y_{n}\dag}_{d_{n+1}}\ket{\bar{\psi}(t_{n+1},\tau)}\| \neq 0$ but $\|\mathcal{L}_{d_{n+1},y_{n+1}}^{Y_n\dag}\ket{\bar{\psi}(t_{n+1},\tau)}\| = 0$ for some $n$, or $p_0(a_0)=0$. Therefore, the sufficient condition for \begin{math}\lambda_{\mathrm{irr}}=0\end{math} is $\|\mathcal{R}^{Y_{n}\dag}_{d_{n+1}}\ket{\bar{\psi}(t_{n+1},\tau)}\|\neq 0$ implies $\|\mathcal{M}_{y_{n+1}}^\dag\mathcal{N}[\mathcal{R}^{Y_{n}\dag}_{d_{n+1}}\ket{\bar{\psi}(t_{n+1},\tau)}]\| \neq 0$, and $p_0(a_0)=0$ for all $a_0$.
We note that we can easily show (via this condition) that the full rank condition is a sufficient condition for \begin{math}\lambda_{\mathrm{irr}}=0\end{math}.

\section{Numerical calculation}\label{Numerical calculation}
In this section, we first provide some additional information about the numerical demonstration in the main text (Fig.~\ref{fig:fluctuation}).
We next show some additional numerical results, which demonstrate the generalized SL and FT under continuous measurement of the imperfect detection rate. We also display the time evolution of the stochastic QC-transfer entropy $i_{\rm QC}$.

\begin{figure}[tb]
\begin{center}
\includegraphics[width=0.45\textwidth]{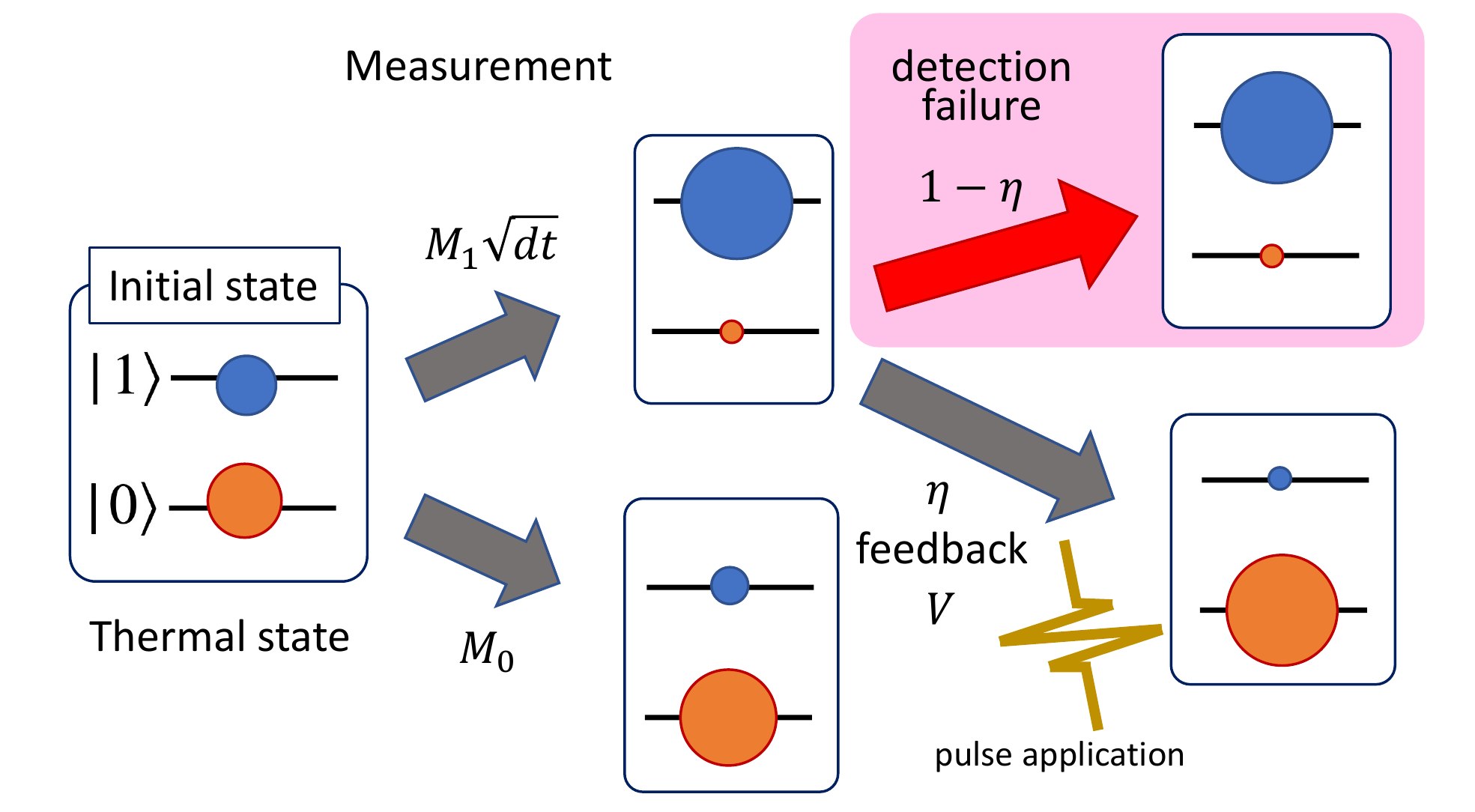}
\caption{A graphical illustration of the setup employed in the numerical calculation for Figs.~\ref{fig:fluctuation}, \ref{fig:generalized FTeta}, \ref{fig:generalized SL}, \ref{fig:iqc}: We reduce the system entropy by increasing the ground state population. Detection failure (\begin{math}\eta <1\end{math}) leads to population increase in the excited state, which prevents the entropy reduction. We set \begin{math}\delta>0\end{math} in the measurement operator \begin{math}M_1\end{math}, so that the state after the measurement $M_1$ is full rank. For simplicity, coherence between \begin{math}\ket{0}\end{math} and \begin{math}\ket{1}\end{math} is neglected in this illustration.}
\label{fig:Numcalc_set}
\end{center}
\end{figure}
The fine unraveled dynamics of the system is calculated by solving SME \eqref{SME_b_t} in the time step \begin{math}\Delta t = 0.01\end{math}. We have checked that the complete degeneracy of \begin{math}\rho_{t_n}^{Y_n}\end{math} does not occur during the no-jump dynamics, which guarantees the validity of Eq.~\eqref{SME_b_t}.
Under measurement with the imperfect detection rate \begin{math}\eta \neq 1\end{math}, the feedback pulse \begin{math}V^{Y_n}\end{math} is applied only after the measurement jump detection (i.e., $y_n=1$). The imperfect measurement prevents the entropy reduction by feedback protocol. Figure~\ref{fig:Numcalc_set} roughly shows the setting employed in the numerical calculation under imperfect measurement. 

\begin{figure}[tb]
\begin{center}
\includegraphics[width=0.45\textwidth]{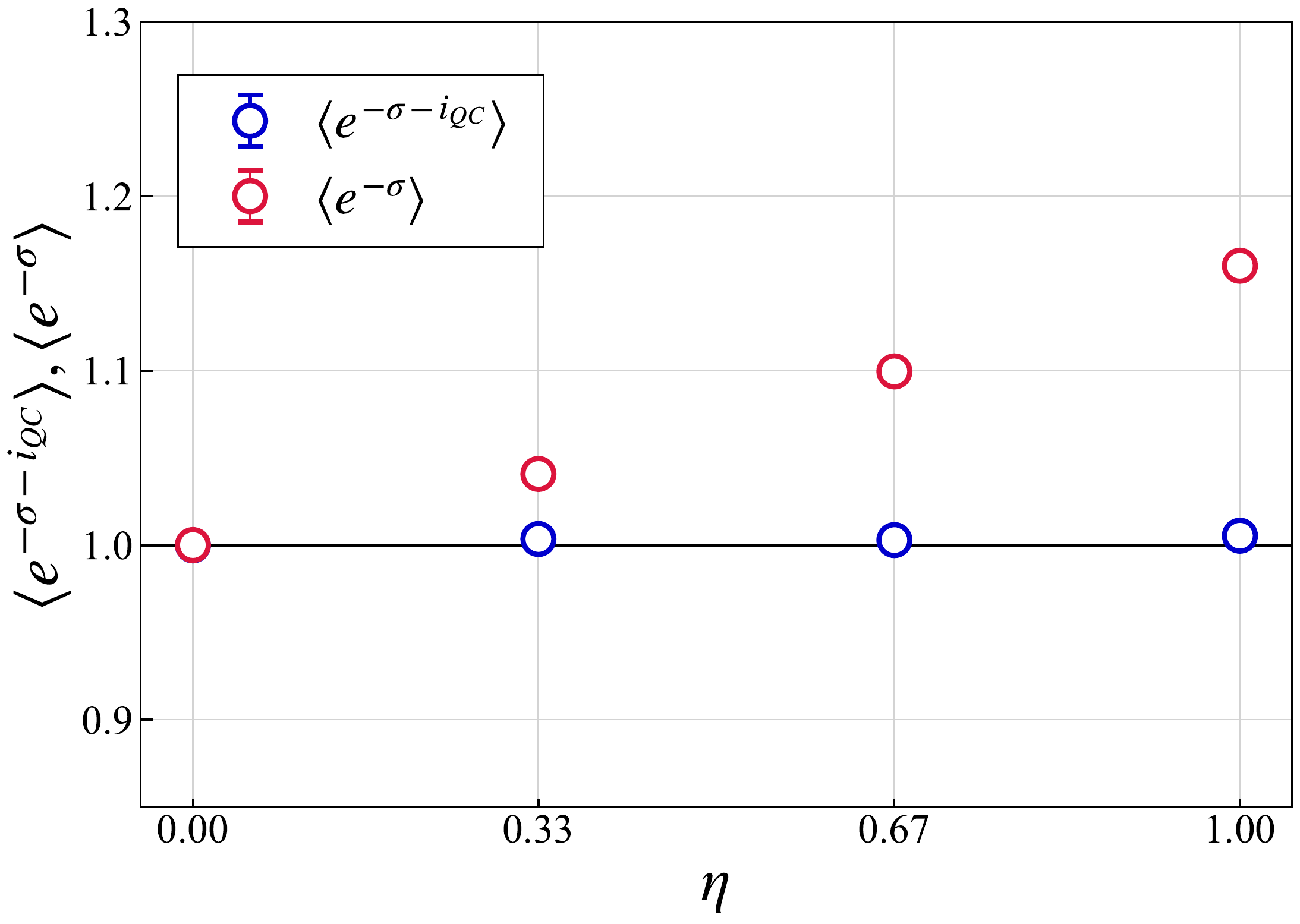}
\caption{Numerical verification of the generalized FT under imperfect measurement: The average values of \begin{math}e^{-\sigma-i_{\mathrm{QC}}}\end{math} and \begin{math}e^{-\sigma}\end{math} are plotted. \begin{math}1.0\times 10^{5}\end{math} trajectories are randomly sampled for each plot. The system parameters are taken to be identical to those in Fig.\ref{fig:fluctuation}. We fixed the trial time \begin{math}\tau=10\end{math} and changed the detection rate as \begin{math}\eta = 0.00,0.33,0.67,1.00\end{math}.}
\label{fig:generalized FTeta}
\end{center}
\end{figure}
Figure~\ref{fig:generalized FTeta} shows the ensemble-averaged values $\langle e^{-\sigma-i_{\rm QC}}\rangle $ and $\langle e^{-\sigma}\rangle$ under continuous measurement with different detection rates.
From Fig.~\ref{fig:generalized FTeta}, we can verify that the generalized FT for imperfect measurement holds for any detection rate. Here, \begin{math}\langle e^{-\sigma} \rangle\end{math} increases as the detection rate gets higher, which reflects that feedback protocol (Fig.~\ref{fig:Numcalc_set}) works better and the system entropy can be reduced more when the detection rate is higher.

\begin{figure}[tb]
\begin{center}
\includegraphics[width=0.8\textwidth]{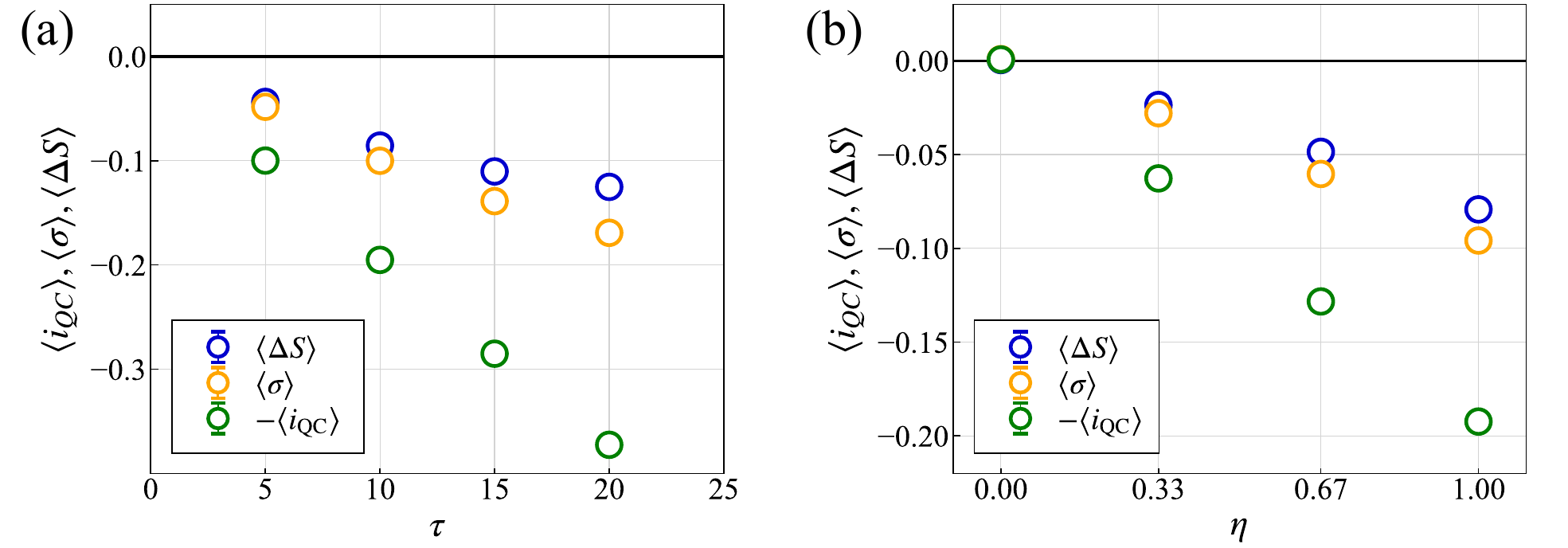}
\caption{A numerical verification of the generalized SL (i.e., $\langle \sigma \rangle \geq -\langle i_{\rm QC}\rangle$) for each (a) trial time and (b) detection rate: The average entropy change \begin{math}\langle\Delta S\rangle \end{math}, the entropy production \begin{math}\langle \sigma\rangle \end{math} and the QC transfer entropy \begin{math}\langle i_{\mathrm{QC}}\rangle \end{math} are plotted. \begin{math}1.0\times 10^5\end{math} trajectories are sampled for each plot. System parameters are the same as those in Fig.\ref{fig:fluctuation}. (a) The detection rate \begin{math}\eta\end{math} is fixed to 1 (perfect measurement) and the trial time is changed as \begin{math}\tau =5,10,15,20\end{math}. (b) The trial time is fixed as \begin{math}\tau =10\end{math} and the detection rate is changed as \begin{math}\eta = 0.00,0.33,0.67,1.00\end{math}.}
\label{fig:generalized SL}
\end{center}
\end{figure}
Figure~\ref{fig:generalized SL} shows that the generalized SL \begin{math}\langle \sigma\rangle \geq -\langle i_{\mathrm{QC}}\rangle\end{math} holds at any time \begin{math}\tau\end{math} and any detection rate \begin{math}\eta\end{math}, though the conventional SL \begin{math}\langle \sigma\rangle \geq 0\end{math} is violated except for \begin{math}\eta =0\end{math}. It can be seen from the decreasing behavior of \begin{math}\langle \Delta S\rangle\end{math} in Fig.~\ref{fig:generalized SL}(a) that the system entropy reduction is achieved at \begin{math}\eta =1\end{math}. The reduction gets less significant as the detection rate decrease, as we can see from Fig.~\ref{fig:generalized SL}(b). It can also be seen that the QC-transfer entropy increases as the time elapses, and also when the detection rate is higher.

\begin{figure}[tb]
\begin{center}
\includegraphics[width=0.8\textwidth]{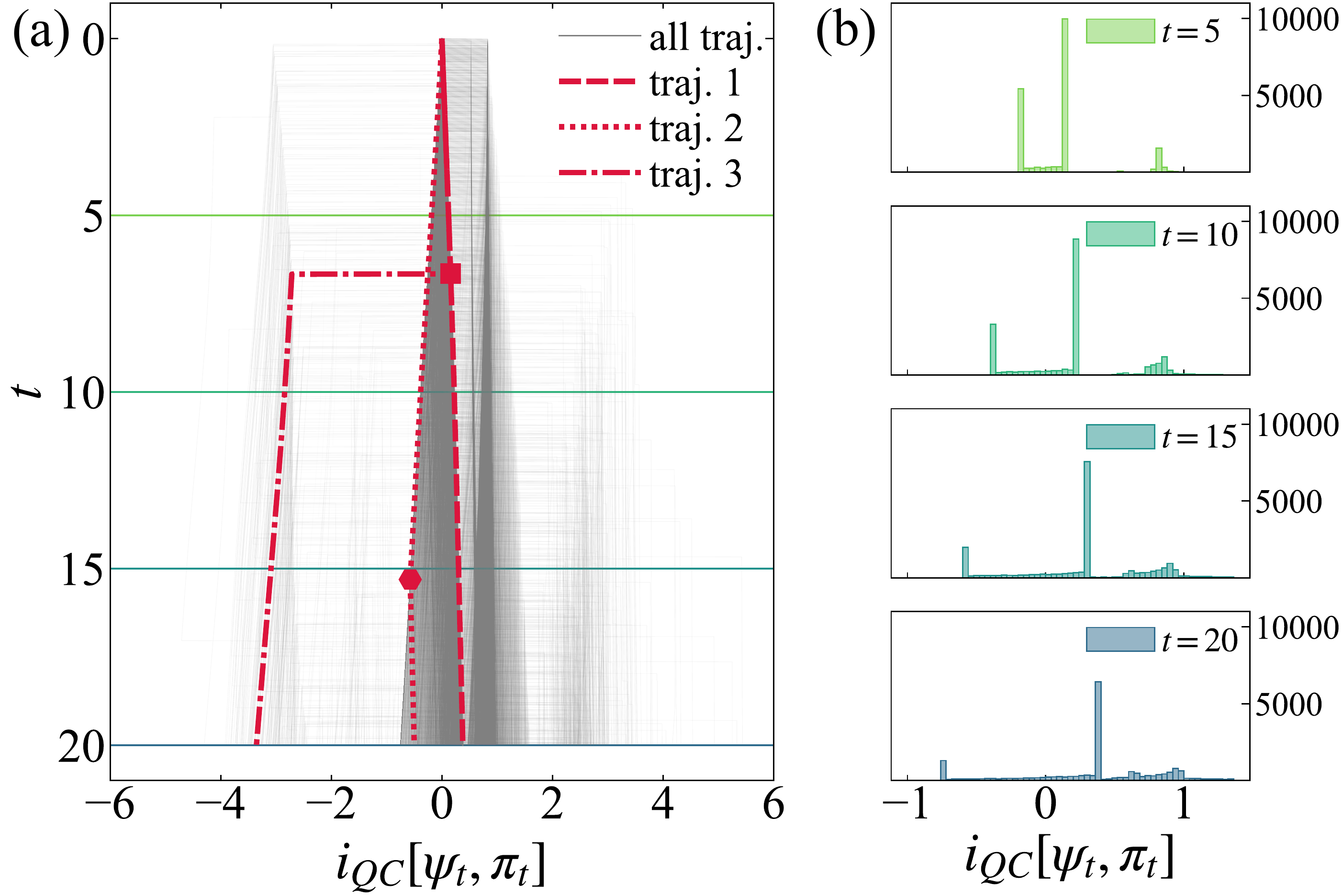}
\caption{(a) Time evolution of the stochastic QC-transfer entropy \begin{math}i_{\mathrm{QC}}[\psi_{t_n},\pi_{t_n}]\end{math}, which are plotted by gray curves. The stochastic QC-transfer entropy \begin{math}i_{\mathrm{QC}}[\psi_{t_n},\pi_{t_n}]\end{math} \begin{math}2.0\times 10^4\end{math} fine trajectories are sampled under the same parameters as those employed in Fig~\ref{fig:fluctuation} of the main text. For better visibility, the transparency of each curve is tuned. We display the three representative trajectories superimposed on these gray curves. Trajectory 1 experiences no quantum jump, trajectory 2 undergoes \begin{math}L_{-}\end{math} jump at the red hexagon, and trajectory 3 goes through \begin{math}M_1\end{math} jump at the red square.
(b)The distribution of \begin{math}i_{\mathrm{QC}}[\psi_{t_n},\pi_{t_n}]\end{math} at \begin{math}t_n=5,10,15,20\end{math}. Histograms of \begin{math}i_{\mathrm{QC}}[\psi_{t_n},\pi_{t_n}]\end{math} at fixed time (shown in (a) as the horizontal line) are lined up vertically. The bin width is 0.04.}
\label{fig:iqc}
\end{center}
\end{figure}
We finally zoom into the stochastic time evolution of \begin{math}i_{\mathrm{QC}}[\psi_{t_n},\pi_{t_n}]\end{math} in Fig.~\ref{fig:iqc}. 
Here, \begin{math}i_{\mathrm{QC}}[\psi_{t_n},\pi_{t_n}]\end{math} denotes the stochastic QC-transfer entropy obtained by the measurement until $t_n$, and defined as 
\begin{equation}
\label{QC_transfer_until_t}
    i_{\mathrm{QC}}[\psi_{t_n},\pi_{t_n}]\equiv \sum_{m=0}^{n-1} -\ln p^{Y_m}(b_{m+1}) +\ln p^{Y_{m+1}}(c_{m+1}).
\end{equation}
As can be seen from Eq.~\eqref{QC_transfer_until_t}, \begin{math}i_{\mathrm{QC}}\end{math} only depends on the measurement results \begin{math}Y_n\end{math} and the repeated PMs \begin{math}\pi_{t_n}\end{math}. 
Since the time evolution of the fine trajectory is uniquely determined as long as quantum jump does not occur, as explained in Section \ref{Time evolution of fine trajectory}, the value of \begin{math}i_{\mathrm{QC}}\end{math} can be uniquely determined until the first quantum jump.
Trajectory 1 in Fig.~\ref{fig:iqc}(a) represents the smooth evolution of no-jump trajectory whose initial state is the ground state of \begin{math}H_t\end{math}. Two dominant peaks in Fig.~\ref{fig:iqc}(b) show the no-jump trajectories started from the ground and the excited state of \begin{math}H_t\end{math} at \begin{math}t=0\end{math}, respectively. The sequence of the histograms reflects the intuition that the number of no-jump trajectories shall decrease exponentially as time elapses.

Let us further focus on the non-smooth changes in $i_{\rm QC}$ caused by quantum jumps. Such changes can be both continuous or discontinuous, depending on whether the source of the jump is heat-bath dissipation or continuous measurement.
Trajectory 2 of Fig.~\ref{fig:iqc}(a) starts from the excited state of \begin{math}H_t\end{math} and experiences dissipation jump \begin{math}L_{-}\end{math} at \begin{math}t_n=15.32\end{math}. Therefore, \begin{math}i_{\mathrm{QC}}\end{math} undergoes non-smooth but continuous change due to the change in \begin{math}(b_n,c_n) \to (b_{n+1},c_{n+1})\end{math}. The bins between two dominant peaks in Fig.~\ref{fig:iqc}(b) represent such trajectories.
On the other hand, when the measurement jump \begin{math}M_1\end{math} occurs, \begin{math}i_{\mathrm{QC}}\end{math} changes discontinuously as shown in trajectory 3 of Fig.~\ref{fig:iqc}(a). The increment of \begin{math}i_{\mathrm{QC}}\end{math} in \begin{math}[t_n,t_{n+1})\end{math} is \begin{math}\mathcal{O}(1)\end{math} (not \begin{math}\mathcal{O}(\Delta t)\end{math}) at the point, because of the abrupt change \begin{math}\rho_{t_n}^{Y_n}\to\sigma_{t_n}^{Y_n,y_{n+1}}\end{math} caused by the measurement jump. 
All the trajectories outside the two dominant peaks in Fig.~\ref{fig:iqc}(b) experience the discontinuous changes caused by measurement jumps. 
In general, the number of quantum jumps tends to increase as time elapses, and consequently the distribution of \begin{math}i_{\mathrm{QC}}\end{math} becomes wider.

\end{document}